\documentclass{aa}  

\usepackage{graphicx}
\usepackage{txfonts}
\usepackage{caption}
\begin{document}

   \title{Mapping the stellar age of the Milky Way bulge with the VVV. 
   \thanks{Based on observations 
taken within the ESO VISTA Public Survey VVV, Programme ID 179.B-2002 (PI: Minniti, Lucas) and VIRCAM@VISTA/ESO, VVV, Programme ID 095.B-0368(A) (PI: Valenti)}}

   \subtitle{ I. The method}

  \author{F. Surot,
          \inst{1,4}
    E. Valenti,
         \inst{1}
    S. L. Hidalgo,
         \inst{2,3}
    M. Zoccali,
         \inst{4,5}
    E. S{\"o}kmen,
         \inst{2,3}
    M. Rejkuba,
          \inst{1,6}
   D. Minniti,
         \inst{5,7,8}    
    O. A. Gonzalez,
          \inst{9}
     S. Cassisi,
         \inst{10,11}
     A. Renzini,
          \inst{12}
      A. Weiss,
         \inst{13}
          }
          
   \institute{European Southern Observatory, Karl Schwarzschild\--Stra\ss e 2, D\--85748 Garching 
bei M\"{u}nchen, Germany. \\
              \email{frsurot@uc.cl}
\and
Instituto de Astrof\'{i}sica de Canarias, V\'{i}a L\'{a}ctea S/N, E\--38200, La Laguna Tenerife, Spain
\and
Department of Astrophysics, University of La Laguna, E\--38200, La Laguna Tenerife, Canary Islands, Spain
\and
Instituto de Astrof\'{i}sica, Pontificia Universidad Cat\'{o}lica de Chile, Av. Vicu\~{n}a Mackenna 4860, Santiago , Chile.
\and
Millennium Institute of Astrophysics, Av. Vicu\~{n}a Mackenna 4860, 782-0436 Macul, Santiago, Chile.
\and
 Excellence Cluster Universe, Boltzmann\--Stra\ss e 2,  D\--85748 Garching bei M\"{u}nchen, Germany
\and
Departamento de Ciencias F\'{i}sicas, Universidad Andr\'{e}s Bello, Rep\'{u}blica 220, Santiago, Chile
\and
Vatican Observatory, V00120 Vatican City State, Italy
\and
Institute for Astronomy, University of Edinburgh, Royal Observatory, Edinburgh EH9 3HJ
\and
INAF \-- Astronomical Observatory of Abruzzo, Via M. Maggini sn 64100 Teramo, Italy
\and
INFN, Sezione di Pisa, Largo Pontecorvo 3, 56127 Pisa, Italy
\and
INAF \-- Osservatorio Astronomico di Padova, Vicolo dell'Osservatorio 5, 35122 Padova, Italy
\and
Max\--Planck\--Institut f{\"u}r Astrophysik, Karl Schwarzschild\--Stra\ss e 1, D\--85748 Garching 
bei M\"{u}nchen, Germany. 
             }

   \date{}

 
  \abstract
   {Recent observational programmes are providing a global view of the Milky Way bulge that serves as template for detailed comparison with models and extragalactic bulges. 
A number of  surveys (i.e. VVV, GIBS, GES, ARGOS, BRAVA, APOGEE) are producing comprehensive and detailed extinction, metallicity, kinematics and stellar density maps of the Galactic bulge with unprecedented accuracy. However, the still missing key ingredient is the distribution of stellar ages across the bulge.}
   {To overcome this limitation, we aim to age\--date the stellar population in several bulge fields with the ultimate goal of deriving an age map of the Bulge. This paper presents the methodology and the first results obtained for a field along the Bulge minor axis, at $b=-6^\circ$. }
   {We use a new PSF\--fitting photometry of the VISTA Variables in the V\'{i}a L\'{a}ctea (VVV) survey data to construct deep color\--magnitude diagrams of the bulge stellar population down to $\sim$\,2\,mag below the Main Sequence turnoff. 
   To address the contamination by foreground disk stars we adopt a statistical approach by using control\--disk fields located at different latitudes \-- spanning approximately the bulge's range \-- and longitudes $-30^{\circ}$ and $+20^{\circ}$.
  We generate synthetic photometric catalogs, including the observational errors and completeness, of complex stellar populations with different age and metallicity distributions. 
   The comparison between the color\--magnitude diagrams of synthetic and observed disk\--decontaminated bulge populations provides constrains on the stellar ages in the observed field.}
   {We find the bulk of the bulge stellar population in the observed field along the minor axis to be at least older than $\sim$\,7.5\,Gyr.
 In particular, when the metallicity distribution function spectroscopically derived by GIBS is used, the best fit to the data is obtained with a combination of synthetic populations with ages in between $\sim$\,7.5\,Gyr and 11\,Gyr. However, the fraction of stars younger than $\sim$\,10\,Gyr strongly depends upon the number of Blue Straggler Stars present in the bulge.
Simulations show that the observed color\--magnitude diagram of the bulge in the field along the minor axis is incompatible with the presence of a conspicuous population of intermediate\--age/young (i.e. $\lesssim 5$\,Gyr) stars.
  }
   {}

   \keywords{Galaxy: structure -- Galaxy: bulge } \authorrunning{Surot et al.}  

   \maketitle
%

\section{Introduction}
\label{sec:intro}
The recent photometric (e.g. VVV \-- \citet{minniti_vvv} and OGLE\,IV \-- \citet{ogleIV}) and spectroscopic  surveys (e.g.  BRAVA \-- \citet{bravaIa,bravaIb}, ARGOS \-- \citet{argosII}, GIBS \-- \citet{gibsI}, GES \-- \citet{GESa,GESb}, and APOGEE \-- \citet{apogeeIa,apogeeIb}) have drastically changed our understanding of the Milky Way bulge. 
The emerging picture is much more complicated than we used to believe until about one decade ago based on the stellar properties observed in few small regions, mostly located along the bulge minor axis.

Owing to the systematic and detailed study of the red clump (RC) stellar population across the whole bulge we have finally reached a comprehensive view of the 3D structure of the bulge \citep[see][for a recent review]{3Dreview}. 
It has been established that our bulge hosts a bar with an orientation with respect to the Sun\--Galactic center line of sight of $\sim$\,27$^\circ$ \citep{wegg+13}, and whose near\--side points in the first Galactic quadrant. 
The bar has a boxy/peanut/X\--shape structure in its outer regions \citep{mcwzoc10,nataf+10,saito+11,wegg+13,ness+16}, a typical morphology of bulges formed out of natural evolution of disk galaxies, and as result of disk dynamical instabilities and vertical buckling of the bar \citep[see e.g.][and references therein]{debattista+06}. The RC distribution suggests also the presence of an axisymmetric structure in the innermost $\sim$\,250\,pc \citep{gonzalez+11c,gerhard+12,valenti+16}, and of a long bar with semi-major axis of $\sim$\,4.6\,kpc, which appears to be the natural thin extension of the main bar at larger radii \citep{wegg+15}.

The spectroscopic metallicity distribution function (MDF) observed in different fields across the bulge is quite broad and complex \citep[i.e. $-1.2\lesssim \mathrm{[Fe/H]} \lesssim +0.7$ dex;][and references therein]{gibsIII}.
It is best represented by two components, metal\--rich (MR)  and metal\--poor (MP), whose relative fraction changes along different lines of sight. 
In particular, MP stars dominate the outer regions, and are characterized by a {\it spherical} spatial distribution. 
On the other hand, the MR population becomes progressively more abundant towards the plane, showing instead a remarkably boxy distribution.

Although spectroscopic surveys of M\--giants (i.e. BRAVA) and RC stars (i.e. ARGOS, GIBS, GES) have shown that, overall, the bulge rotates cylindrically like a bar \citep{bravaIa,kunder+12,ness+13ArgosKin,gibsI,ness+16ApogeeKin}, there are compelling observational evidences that demonstrate that MR and MP stars have different kinematics \citep{babusiaux+10, kunder+16,zasowski,gibsIII, rojas-arriagada+17,clarksonmet}.  
MR stars show a steep velocity dispersion gradient as a function of the latitude, from $\sigma\sim 50$\,km/s at $b=-8^\circ$ up to $\sigma\sim140$\,km/s at $b=-1^\circ$. 
Instead, the MP component has a dispersion that ranges from $\sim$\,80\,km/s in the outer region to $\sim$\,120\,km/s at $b=-1^\circ$.

While there is a general agreement regarding the properties of the bulge morphology (i.e. 3D structure), metallicity and kinematics, an unanimous consensus on the age is still missing. 
Indeed, in the recent years the age of the bulge stellar population has been the most controversial problem because of a number of contradicting results based on different approaches.

Dating bulge stars is a very complicated task, challenged by the stellar crowding, the large and high differential extinction, the uncertainties in the distance modulus, the distance spread due to the spatial depth of the bulge/bar along the line of sight, the metallicity dispersion, and finally the contamination by foreground disk stars. The different contribution of all these factors prevents accurate location in terms of magnitude and color of the main sequence turnoff (MS\--TO) of the bulge population, so far among the most reliable age diagnostics \citep{renzini+88}.
Historically, the earliest age constraint by \citet{vandenbergh+74} in the Plaut field along the bulge minor axis at $b=-8^\circ$ ($\sim$1\,kpc) indicated a globular cluster\--like age. \citet{terndrup88} fitted the photometry of bulge fields at a range of latitudes with globular cluster isochrones of varying metallicity, but because lacking a secure distance for the bulge only a weak age constraint (11\--14\,Gyr) was derived. 
\citet{ortolani+95} solved the problem of the distance uncertainties by comparing the bulge population with the two clusters NGC\,6528 and NGC\,6553. 
Specifically, by measuring the difference between the RC and the MS\--TO magnitude in the bulge field and in the clusters they showed, for the first time, that the relative ages of the bulge and metal\--rich cluster population could not differ by more than 5\%. 
\citet{feltzing+00} used the counts of stars brighter and fainter than the MS\--TO observed in their HST\--based photometry of Baade's window and another low extinction field known as the Sgr\--I (i.e. at $l=1.25^\circ$ and $b=2.65^\circ$) to argue in favor of an old age. 
The case for an {\it purely old }bulge has been further strengthened by later studies based on more accurate HST and ground based photometry of different bulge fields located mostly along the minor axis \citep{kuijken+02,clarkson08PPM,clarkson11BS,zoccali+03}, but also at the edge of the bar \citep{valenti+13}. 
Unlike previous works, the problem of contamination from foreground disk stars was tackled either kinematically by using proper motions \citep{kuijken+02,clarkson08PPM,clarkson11BS,calamidaimf,calamidawd}, or statistically by considering control disk fields \citep{zoccali+03, valenti+13}.  
From the analysis of optical and near\--IR decontaminated color\--magnitude diagrams (CMD), these studies found the bulk of the bulge stellar population to be old (i.e. $>$\,10\,Gyr), with no evidence of significant age differences between the field and old Milky Way (MW) cluster population.  
In particular, \citet{clarkson11BS} provided an upper limit of $\sim$\,3.4\% for a bulge component younger than 5\,Gyr, although arguing that the majority of the stars brighter than the old MS\--TO in their CMD could be blue straggler stars (BSS). 

There is, however, a clear discrepancy between the ages inferred from the determination of the MS\--TO in the observed CMDs and those derived by the microlensing project of Bensby and collaborators  \citep{bensby+13, bensby+17}, which derives individual stellar ages from the effective temperature and gravity (i.e. from isochrones in the [$T_{eff}$, log\,g] plane) as obtained from high resolution spectra.
Based on a sample of 90 F and G dwarf, turnoff and subgiant stars in the bulge  (i.e. $|l|\lesssim6^\circ$ and $-6^\circ< b < 1^\circ$) observed during microlensing, \citet[][hereafter B17]{bensby+17} found that about 35\% of the MR stars ([Fe/H]$>0$) span ages in between 8\,Gyr and 2\,Gyr, whereas the vast majority of MP  ([Fe/H]$\lesssim-0.5$) are 10\,Gyr or older. 
In addition, from the derived age\--metallicity and age\--$\alpha$ elements distribution the authors concluded that the bulge must have experienced several significant star formation episodes, about 3, 6, 8 and 12\,Gyr ago. 
Comparable results have been found by \citet{schultheis+17}, who presented the age distribution of 74 giants in the Baade's Window as a function of stellar metallicity.  
Specifically, the relation of \citet{martig+16} calibrated on asteroseismic data  has been used to link the [C/N] abundances measured from APOGEE spectra to the stellar age.  
While the age distribution of the MP ($\mathrm{[Fe/H]<-0.1}$) giants peaks at 10\,Gyr with a decreasing tail towards younger age (as young as 2\,Gyr), MR ($\mathrm{[Fe/H]>-0.1}$) stars can be either young or old. Indeed, their age distribution appears bimodal, with two peaks at 4 and 11\,Gyr. 

Different concepts have been explored and proposed to partially reconcile the spectroscopic and photometric ages.  In this respect the first attempt was presented by \citet{nataf+12} and \citet{nataf16} who proposed a higher helium enrichment factor than currently adopted for the MR isochrones.  
The use of standard isochrones on He\--enhanced stellar populations would lead to photometric and spectroscopic ages that are over\-- and under\--estimated, respectively. Therefore, the discrepancy could be interpreted under the assumption that the chemical evolution of the bulge is He\--enhanced.
On the other hand, \citet{haywood+16} suggested the discrepancy being caused by the effect of the age\--metallicity degeneracy that makes it hard to distinguish in the CMD a young MR star from an old MP one. 
They compared the MS\--TO color spread observed in the CMD of \citet{clarkson11BS} with that of synthetic CMDs obtained by using two different age\--metallicity relations:  {\it i)} the one presented by \citet{bensby+13}, based on a total sample of 59 microlensed dwarfs, and {\it ii)} one that extends from $\mathrm{[Fe/H]}=-1.35$\,dex at 13.5\,Gyr to $\mathrm{[Fe/H]}=+0.5$\,dex at 10\,Gyr. 
When taking into account distance, reddening and metallicity effects, \citet{haywood+16} showed that the MS\--TO color spread of a {\it purely} old stellar population would be wider than what is observed, and thus advocating for the presence in the bulge of a conspicuous population of young and intermediate\--age stars.
Very similar results have been presented by  \citet{bernard+18} who calculated the star formation history of four bulge fields, including that of  \citet{clarkson11BS}. Their findings suggest that over 80\% of the stars are older than 8\,Gyr, but also the presence of star formation as recent as $\sim$\,1\,Gyr.

Clearly, as of today the age distribution of the bulge is still not universally understood, and in particular its spatial variation across the large area of the bulge has not been fully explored. 
In this framework, the use of near\--IR deep photometry provided by the VVV survey represents a unique opportunity. 
With the ultimate ambitious goal of deriving the age map of the stellar population across the bulge, in this paper we present the methodology used to derive stellar ages from the CMDs and the first results obtained on a field located along the bulge minor axis, at $b=-6^\circ$. 

\section{Observations and data reduction}
\label{sec:data}
 
This analysis makes use of a combination of J and $\mathrm{K_s}$ single\--epoch observations of bulge and disk fields collected with the wide field near\--infrared imager VIRCAM mounted at the VISTA telescope at the ESO Paranal Observatory.  
The bulge data for a field at $(0^\circ, -6^\circ)$ has been taken from the VVV survey. In addition, 8 disk fields  with latitudes in the range $-8^\circ<b<+4^\circ$, and longitude $-30^{\circ}$ and $+20^{\circ}$ (see Table\,\ref{tab:obs-log}) were observed in Service Mode as part of the programme 095.B-0368(A) (PI: Valenti) by using exactly the same VVV observing strategy.

The VISTA/VIRCAM focal plane is a mosaic of 16 detectors with gaps of the order of the size of the detectors between them. A single image is called pawprint, which comprises a single exposure of all 16 detectors, producing an image with gaps. Additionally, a jitter of $\sim$\,60 pixels is applied to subtract sky background. In order to fill the gaps in the individual exposures, a series of 6 pointings are combined in a $2\times3$ dither \textit{pawprint pattern}, which ensures more or less homogeneous coverage in a $\sim\,1.5\times 1.2\,\mathrm{deg}^2$ field of view. This strategy also helps to diminish the effects of defects in particular detectors, by having almost each pixel exposed at least twice.
 
However, the overlap areas between individual pawprints and edges of the tiles have 2\--6 times higher exposures, meaning that the noise distribution within a tile varies strongly with position in the sky. 
This drove the decision to run the photometry on stacked pawprint images, rather than the composite tile, as described below.


\begin{table}
	\caption{Observed bulge and disk fields.}             
	\label{tab:obs-log}      
	\centering          
	\begin{tabular}{ c c c c l l l }     
		\hline\hline       
		Name & $(l, b)$ & $<E(\mathrm{J-K_s})>^a$ & Detected\\ 
		  &  &   & stars \\ 
		\hline                    
b249 & $( -0.45^\circ , -6.39^\circ )$ & $ 0.150  \pm  0.047 $ & 2,728,265 \\
c001 & $( -30.00^\circ , -2.97^\circ )$ & $ - $ & 2,069,471 \\
c002 & $( -30.00^\circ , -5.97^\circ )$ & $ - $ & 1,153,139 \\
c003 & $( -30.00^\circ , -7.98^\circ )$ & $ - $ & 723,786 \\
c004 & $( -30.00^\circ , +3.98^\circ )$ & $ - $ & 1,537,790 \\
c005 & $( +19.99^\circ , -2.97^\circ )$ & $ - $ & 2,555,073 \\
c006 & $( +20.00^\circ , -5.98^\circ )$ & $ - $ & 1,524,750 \\
c007 & $( +19.99^\circ , -7.98^\circ )$ & $ - $ & 1,044,536 \\
c008 & $( +19.99^\circ , +3.97^\circ )$ & $ - $ & 1,711,455 \\
		\hline \hline
	\multicolumn{4}{l}{$^{a}$ Color excess from \citet{oscar12}, average and}\\
	 \multicolumn{4}{l}{ standard deviation are taken over the whole tile.}
	\end{tabular}
\end{table}
\begin{figure}
	\centering
	\includegraphics[width=7 cm]{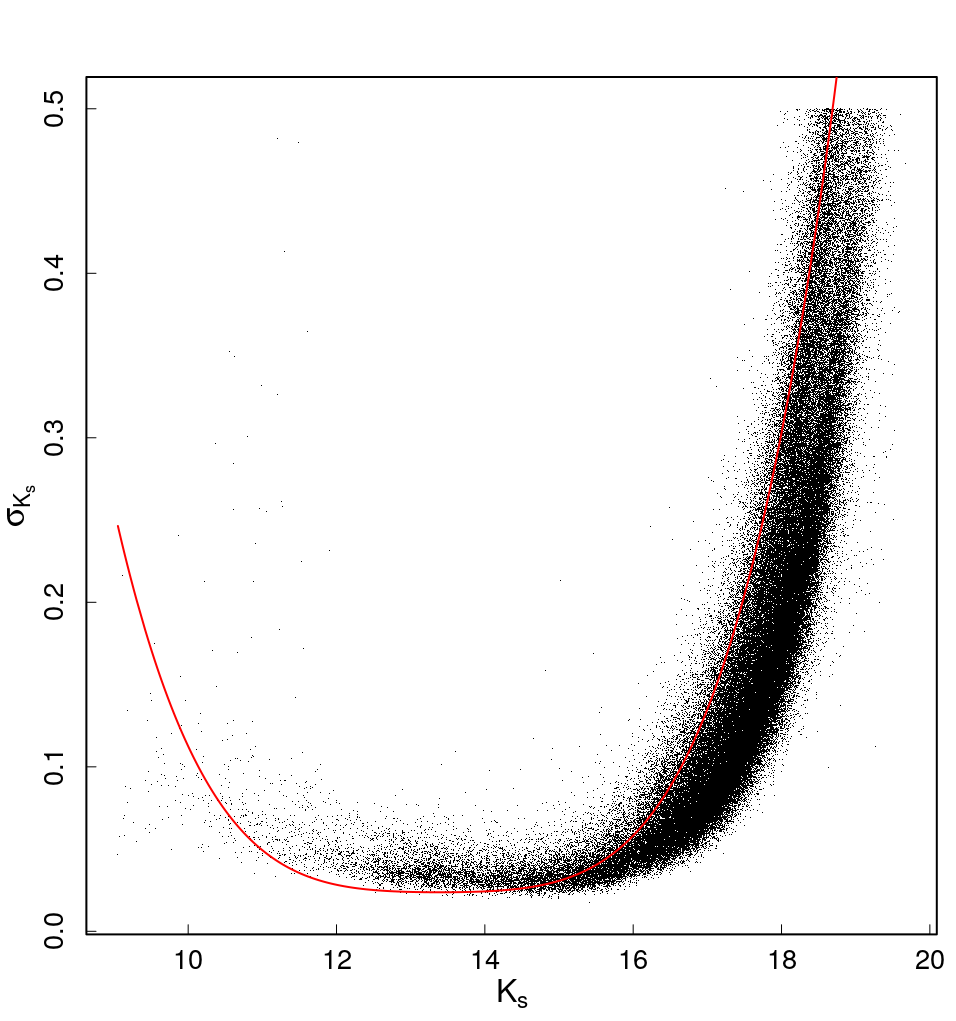}
	\caption{Distribution of photometric errors as a function of magnitude (black dots) in the $\mathrm{K_s}$ band of detector \#12 for the bulge field b249. With the exception of the magnitude range $15 \gtrsim K_s \gtrsim 11$, the photometric errors are well below the dispersion calculated from simulations (red line). Also worth noticing, is the effect from saturation for stars with $K_s \lesssim 12$.}
	\label{fig:error}
\end{figure}   

 All raw data has been processed by the VISTA data flow system pipeline \citep{CasuPipe} at CASU, which among a number of final products\footnote{http://casu.ast.cam.ac.uk/vistasp/} provides the tiled and stacked pawprint images. 

We carried out photometry including PSF modeling on each single detector from the pawprint image by using a customized pipeline based on DAOPHOT \citep{daophot} and ALLFRAME \citep{allframe}. 
 The so\--derived single detector catalogs are combined together using error weighted means in case of overlap to produce single\--band catalog of a tile. Then, for each observed bulge and disk  fields, we derive a photometric catalog listing the instrumental J and $\mathrm{K_s}$ magnitudes through cross\--correlation of the single\--band catalogs. 
Finally, the CASU catalogs have been used to perform the absolute calibration (by means a of simple magnitude shift using several thousands stars in common) and astrometrization onto the VIRCAM photometric and astrometric system. More technical details specific to catalogs cross\--match and internal detector\--by\--detector calibration will be provided in a companion paper \citep{photopaper}, where the release of the PSF\--fitting photometry of the entire VVV bulge area (i.e. $-10^\circ < l < 10^\circ$ and $-10^\circ < b < +5^\circ$) is presented.

We estimated the photometric and systematic errors affecting the derived magnitudes by means of artificial star experiments.
Specifically,  $\sim \!\! 200,000$ artificial stars per detector were added to the stacked pawprint images with random magnitudes J and $\mathrm{K_s}$ compatible with those observed in the instrumental CMDs. In other words, the artificial stars were added along the observed CMD sequences, including their spread. In each independent experiment, the artificial stars were simulated by using the same PSF model obtained during the photometric reduction process, and spatially distributed around a grid properly customized to avoid artificially increasing the crowding \citep[see][for further details]{zoccali+00}.
\begin{figure*}[ht]
\centering
\includegraphics[width=6 cm]{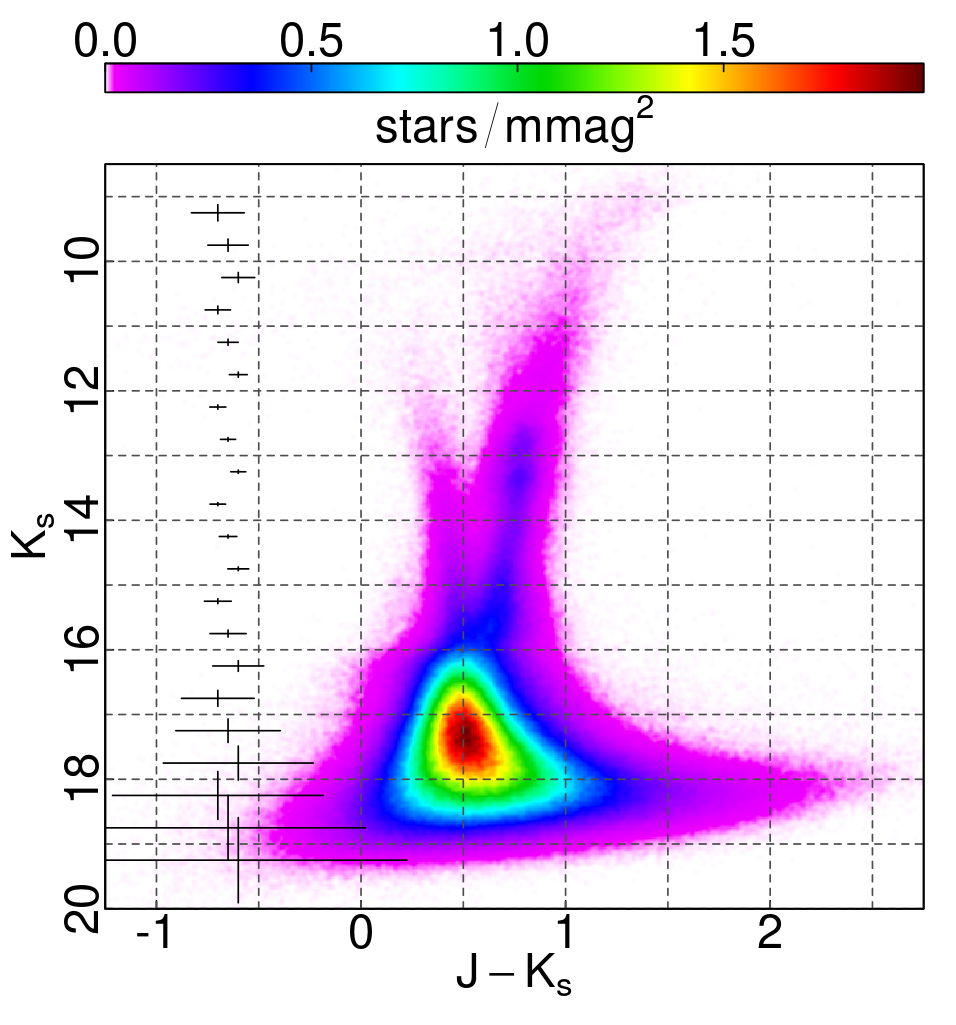}
\includegraphics[width=6 cm]{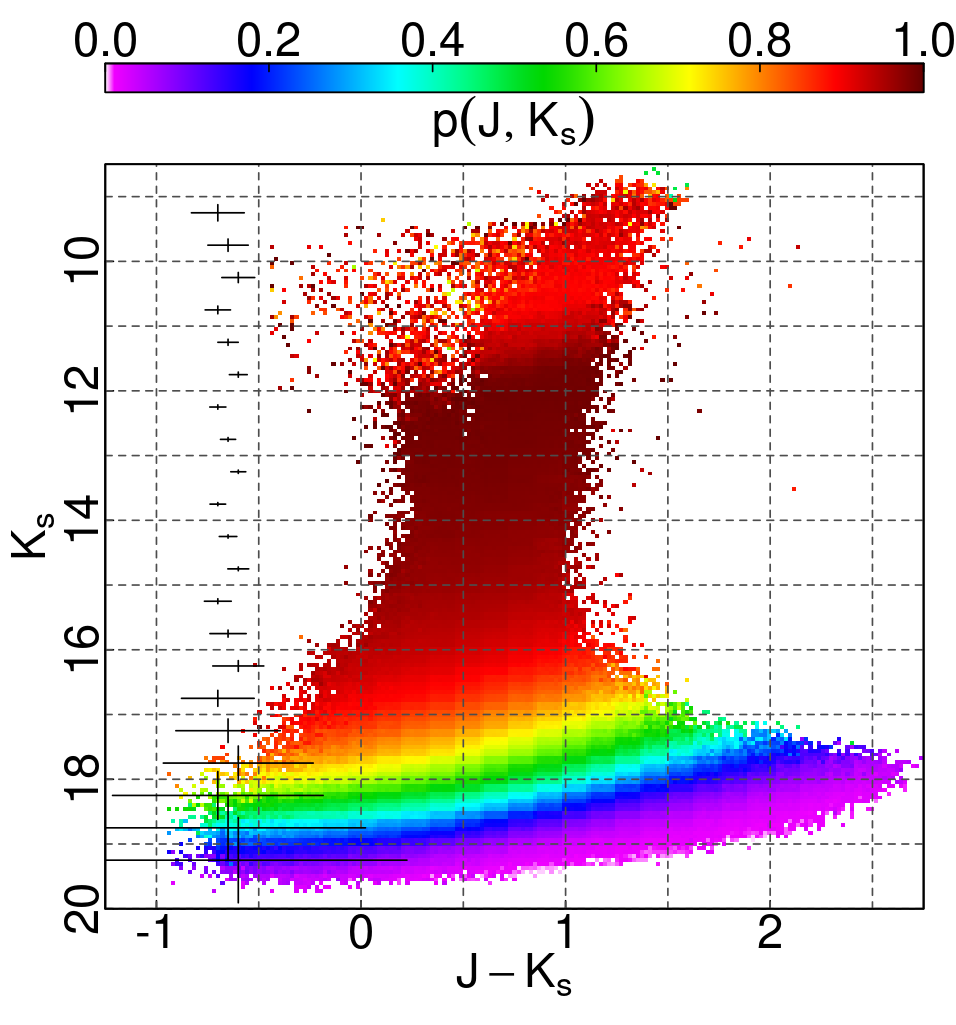}
\includegraphics[width=6 cm]{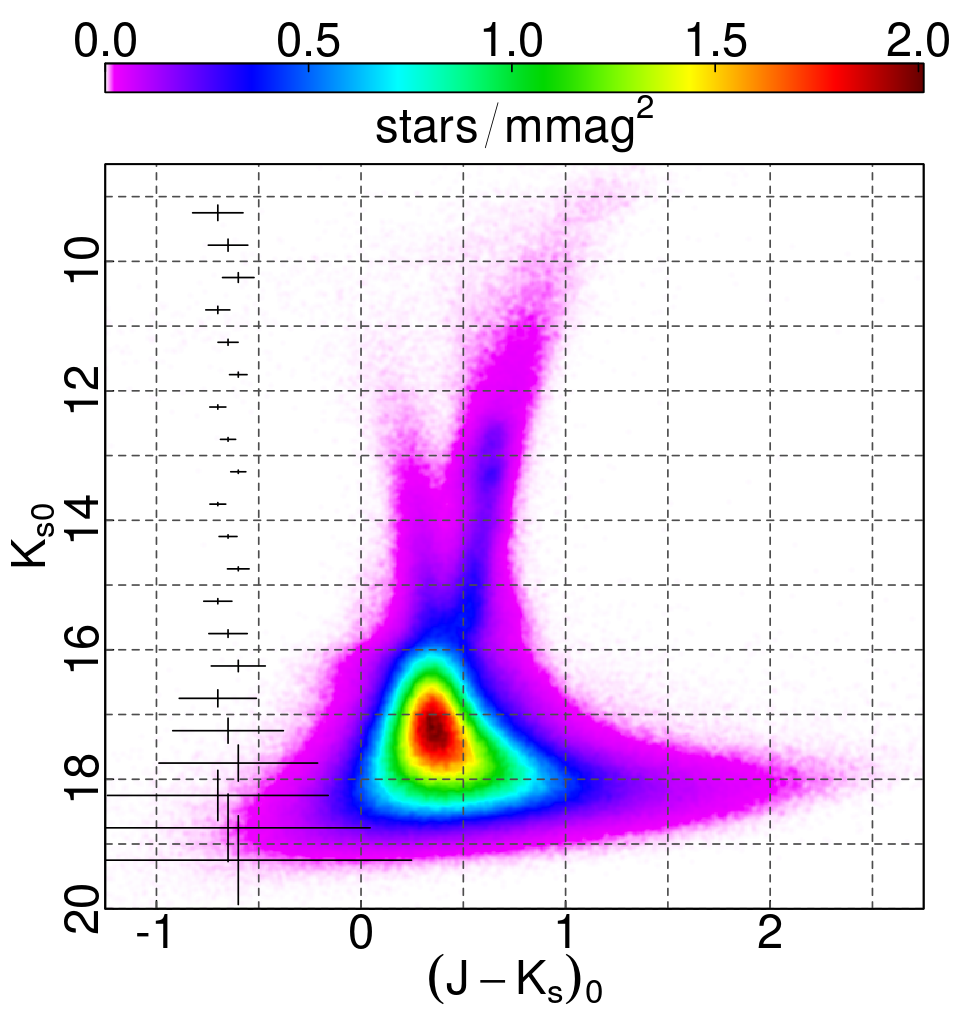}
   \caption{Observed (left) CMDs of the VVV field (b249) shown as a Hess density diagram, with the corresponding photometric completeness map (middle) as detailed in \S\ref{sec:data}. Right\--hand panel shows the reddening\--corrected CMD of the same field. Typical errors on the color and magnitude are also shown as cross symbols for all cases.}
      \label{fig:BulgeObsCMDs}
\end{figure*}

As shown in Fig.~\ref{fig:error}, the combined effect of systematics and photometric uncertainties produces a spread in the recovered vs. injected magnitudes (red solid line) that is considerably larger than what one would expect from the photometric error alone (black dots). It also evidences the known issue related to the saturation of bright stars in the VVV $\mathrm{K_s}$\--band images ($\mathrm{K_s} \sim 12$). Of course there is also saturation in the J\--band, but the uncertainties up to $\mathrm{J} \sim 10$ are well in line with the rest of the profile, from both photometric and simulated sources.

\begin{figure*}
\centering
\includegraphics[height=5 cm, trim={0 3.9cm 5.7cm 0}, clip]{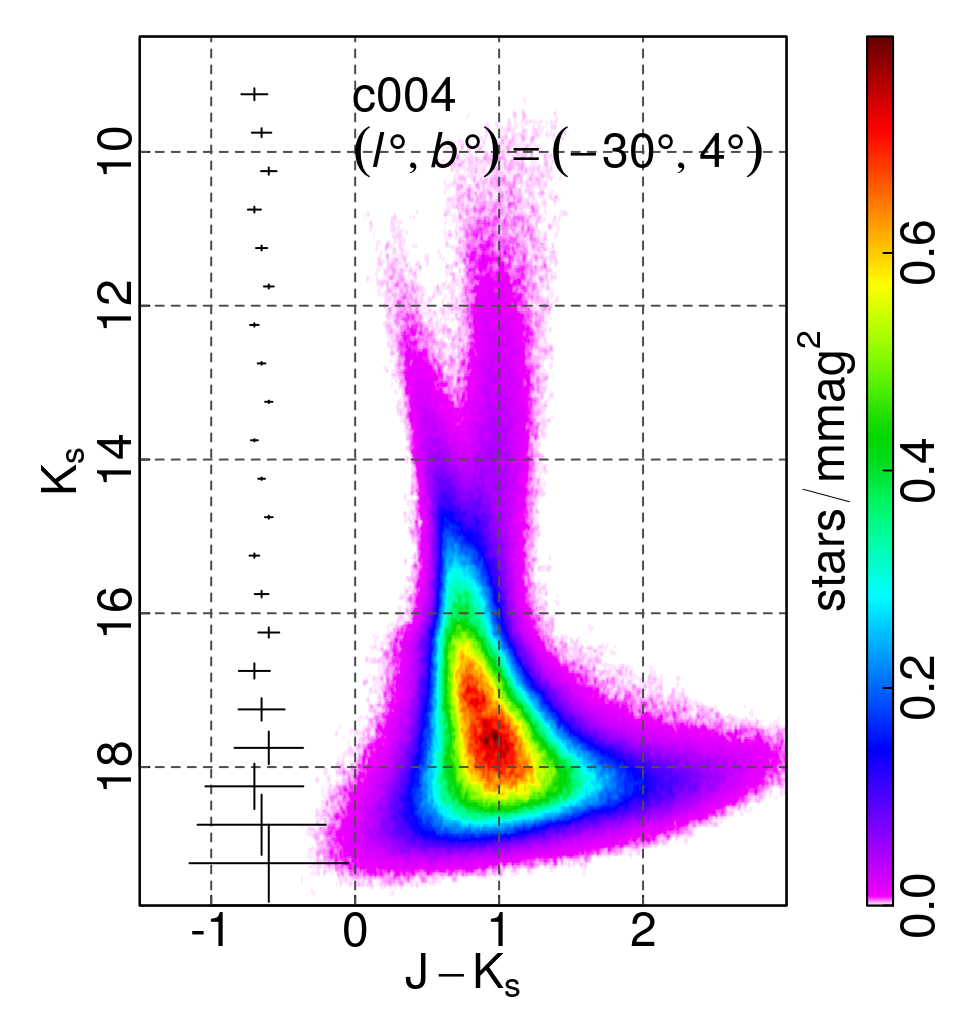}
\includegraphics[height=5 cm, trim={4.8cm 3.9cm 0 0}, clip]{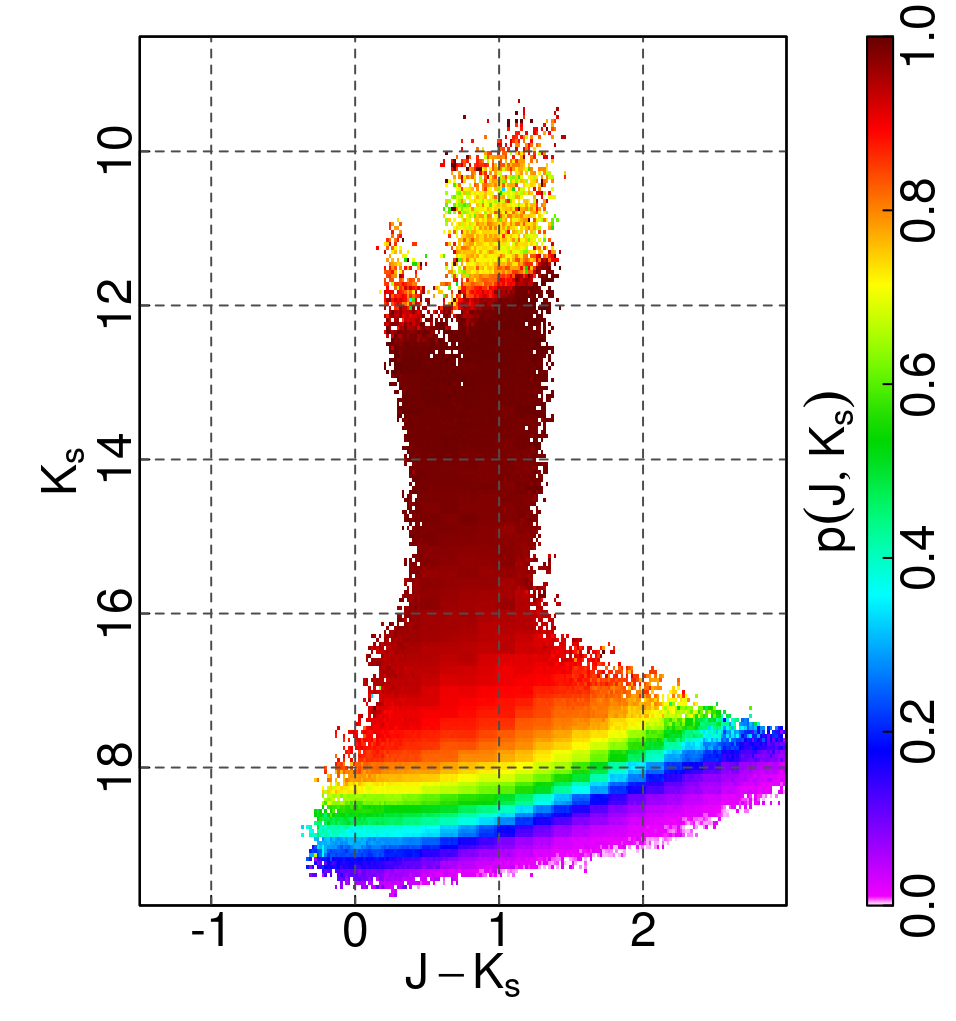}
\includegraphics[height=5 cm, trim={0 3.9cm 5.7cm 0}, clip]{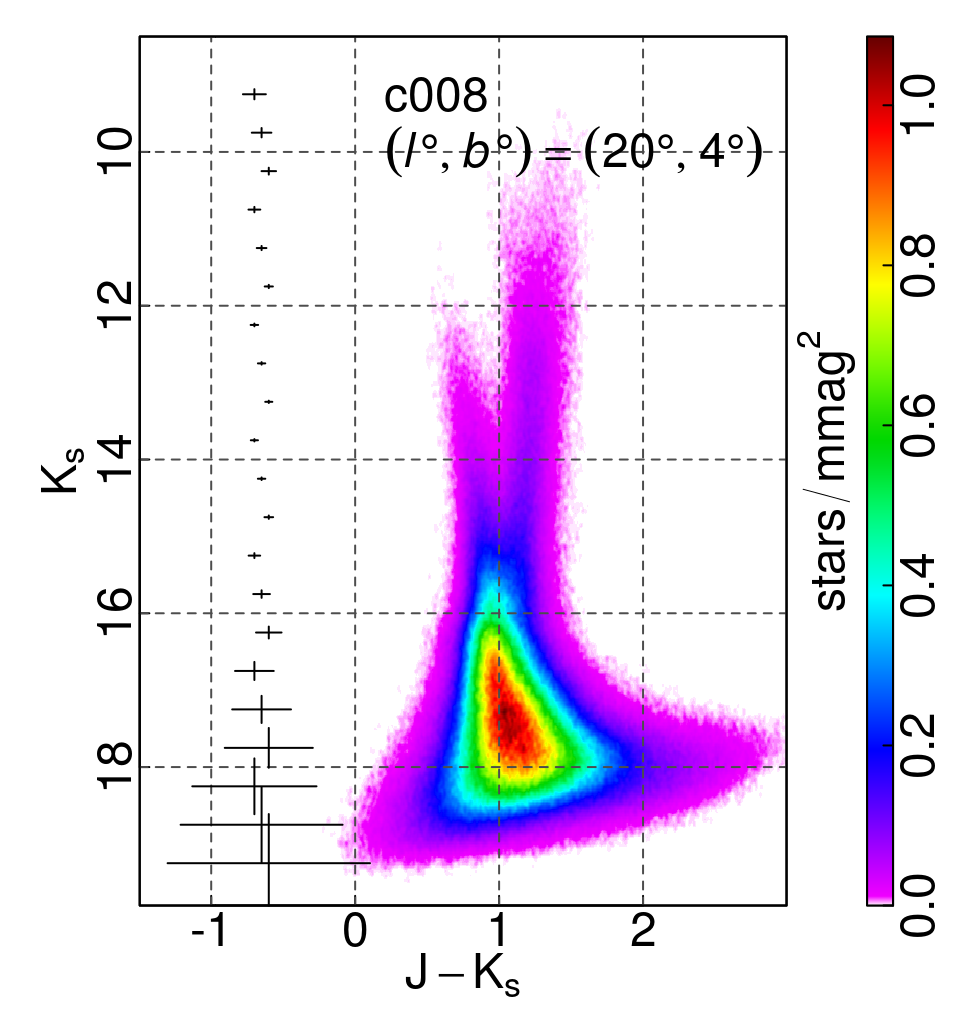}
\includegraphics[height=5 cm, trim={4.8cm 3.9cm 0 0}, clip]{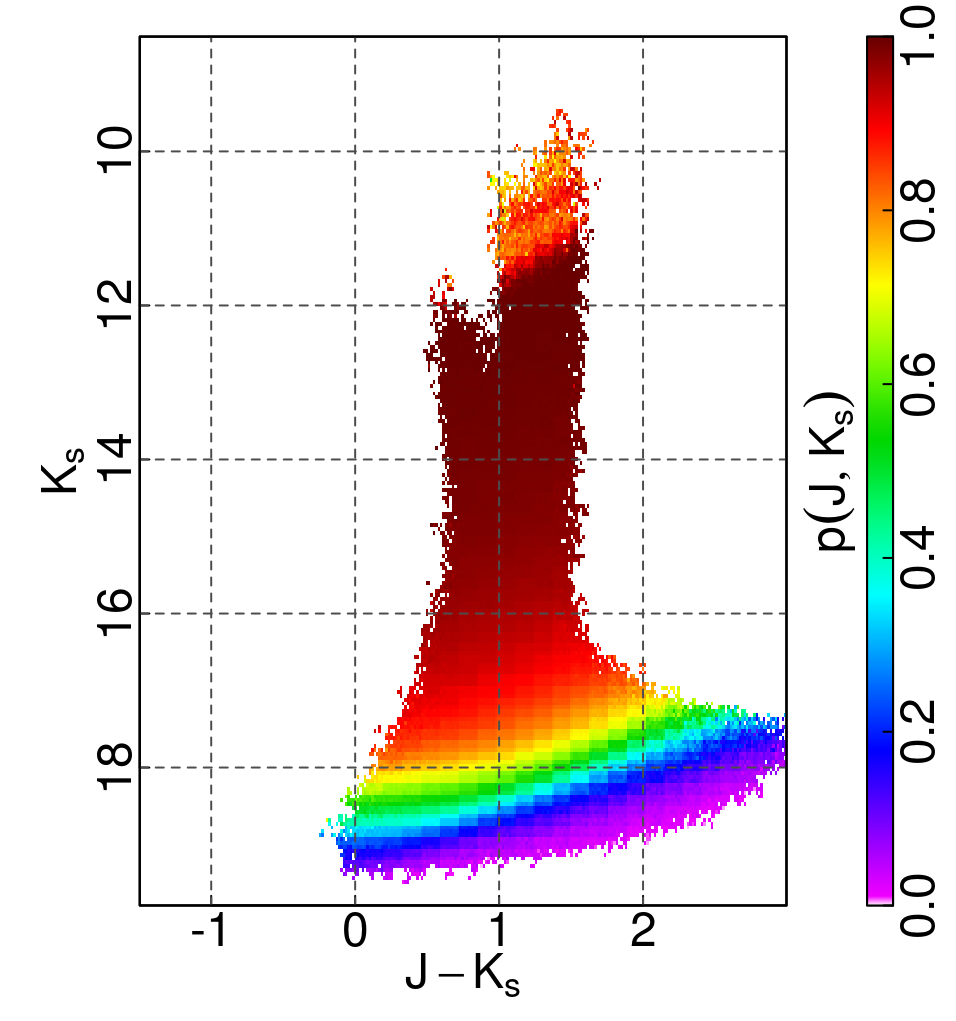}
\includegraphics[height=5 cm, trim={0 3.9cm 5.7cm 0}, clip]{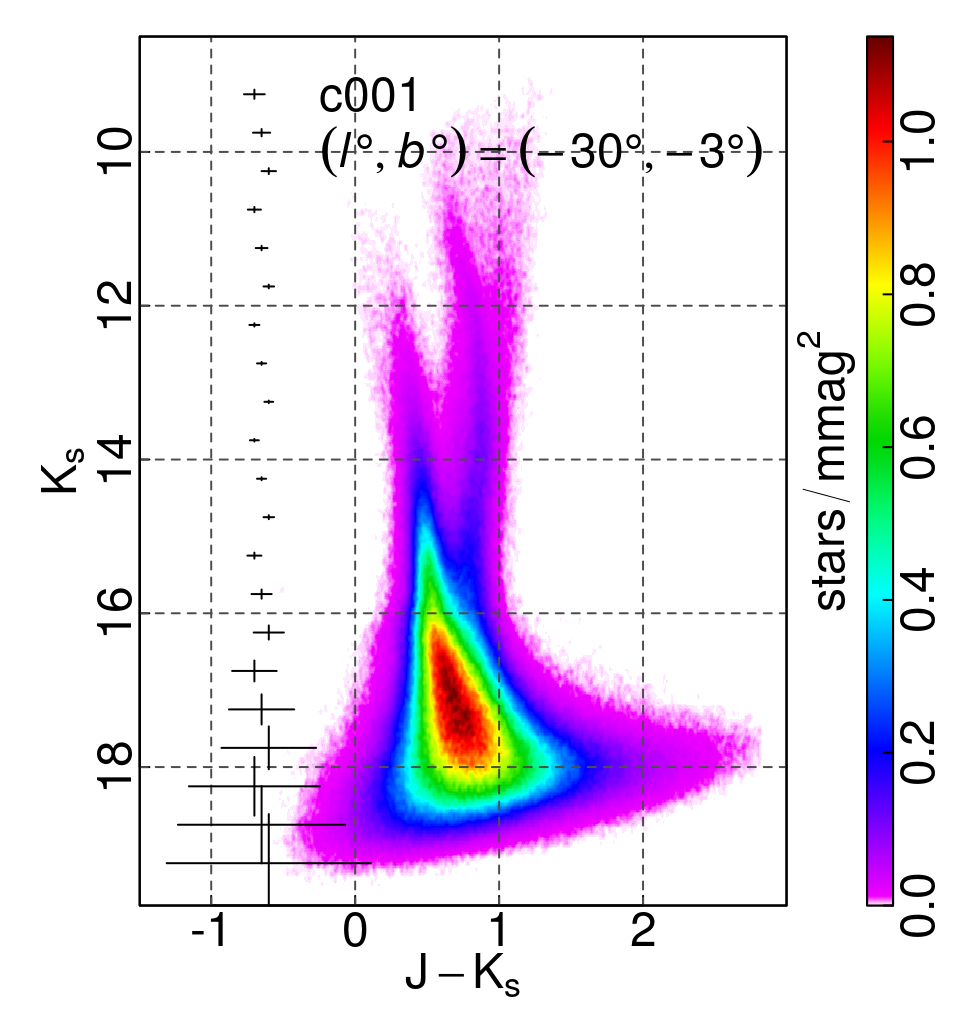}
\includegraphics[height=5 cm, trim={4.8cm 3.9cm 0 0}, clip]{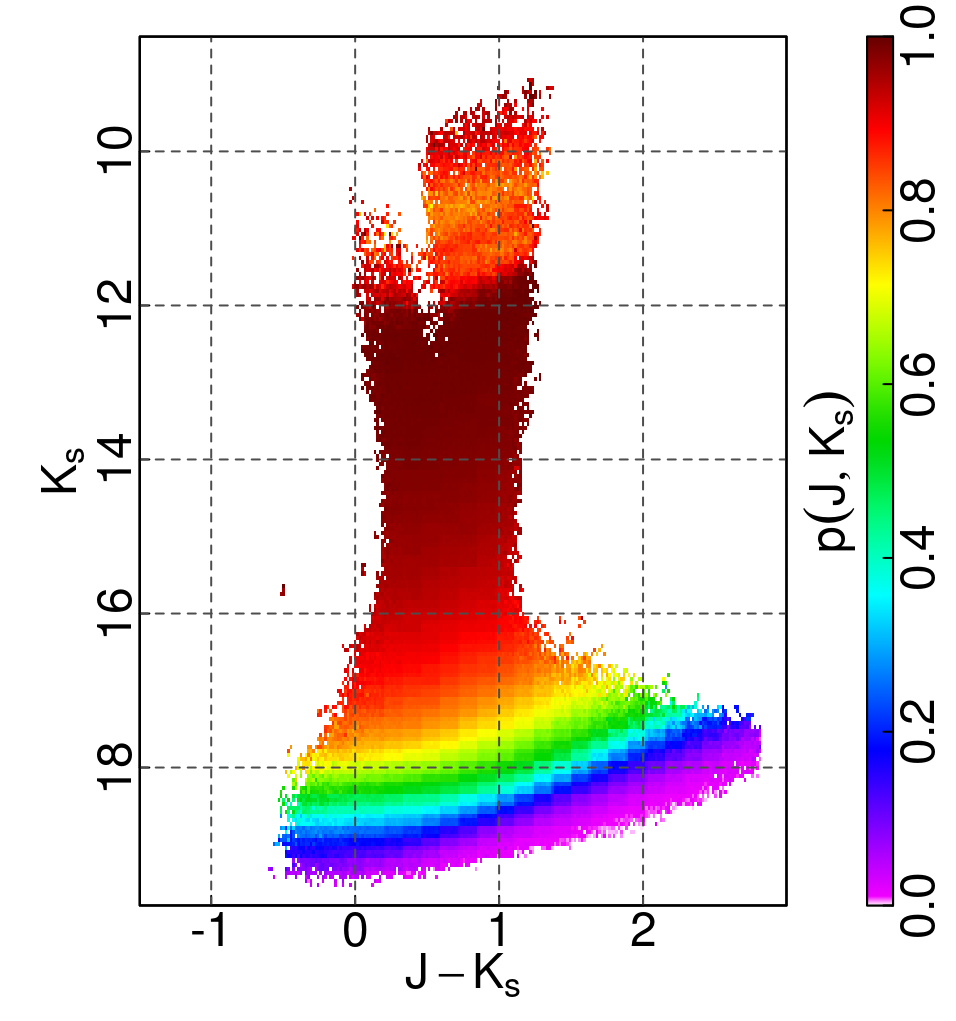}
\includegraphics[height=5 cm, trim={0 3.9cm 5.7cm 0}, clip]{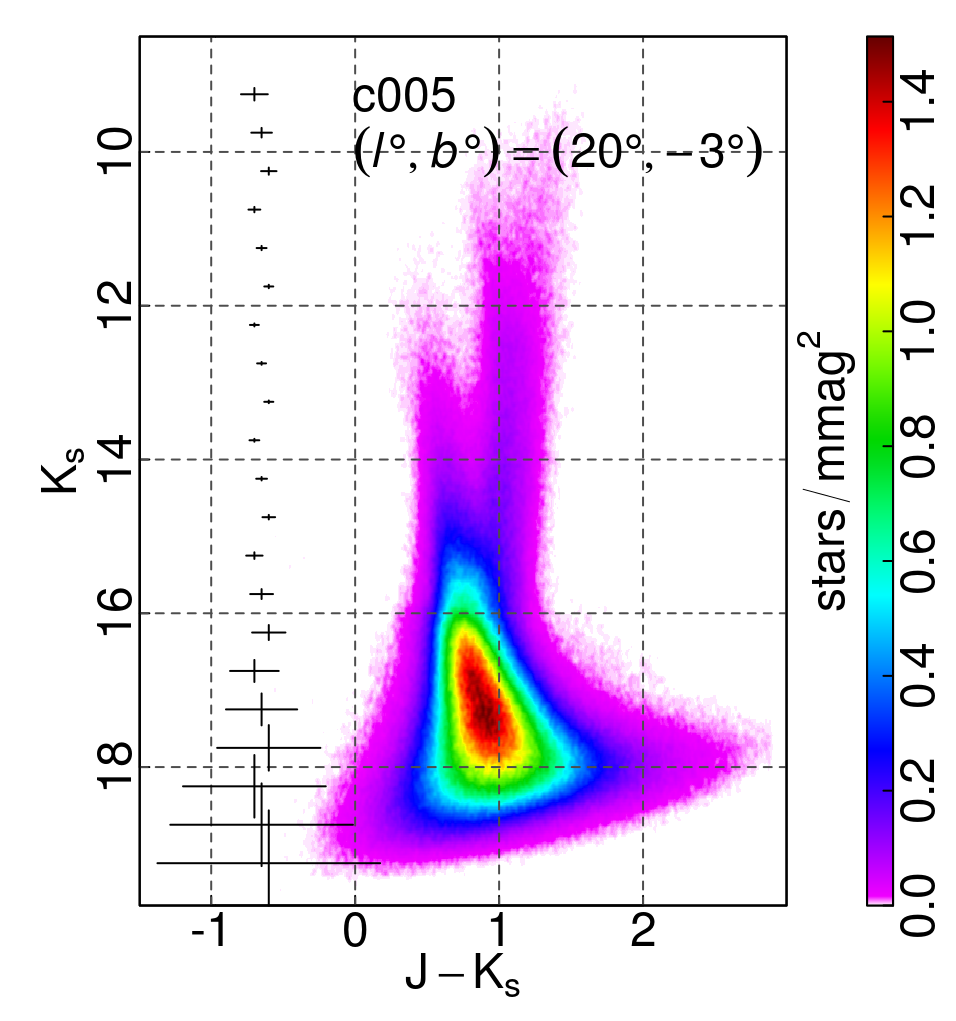}
\includegraphics[height=5 cm, trim={4.8cm 3.9cm 0 0}, clip]{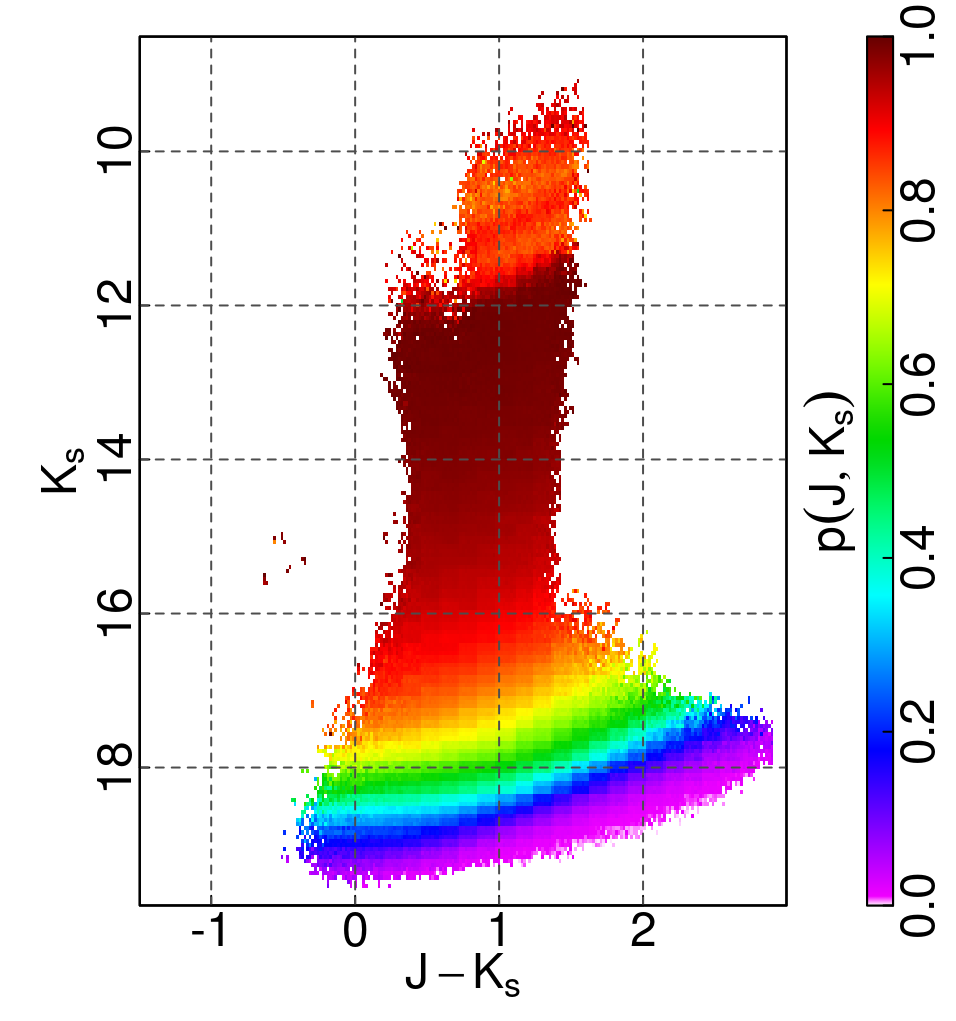}
\includegraphics[height=5 cm, trim={0 3.9cm 5.7cm 0}, clip]{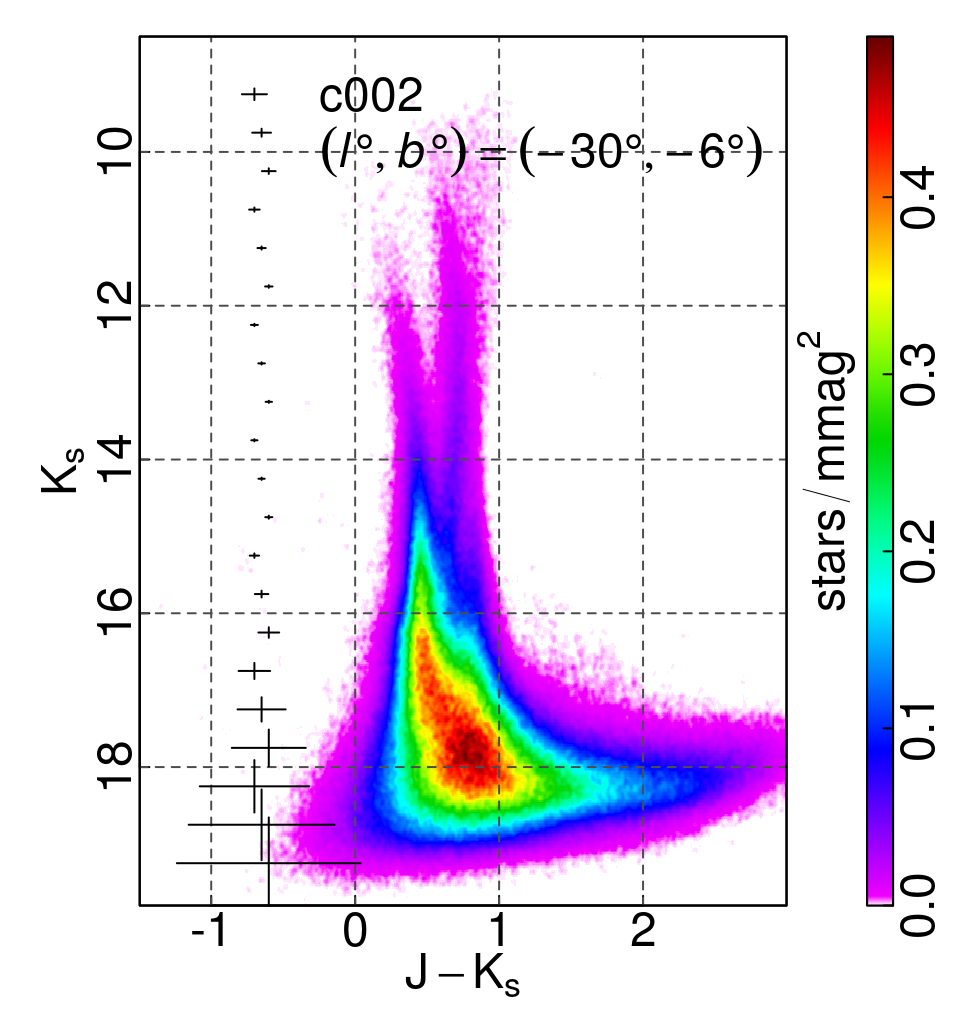}
\includegraphics[height=5 cm, trim={4.8cm 3.9cm 0 0}, clip]{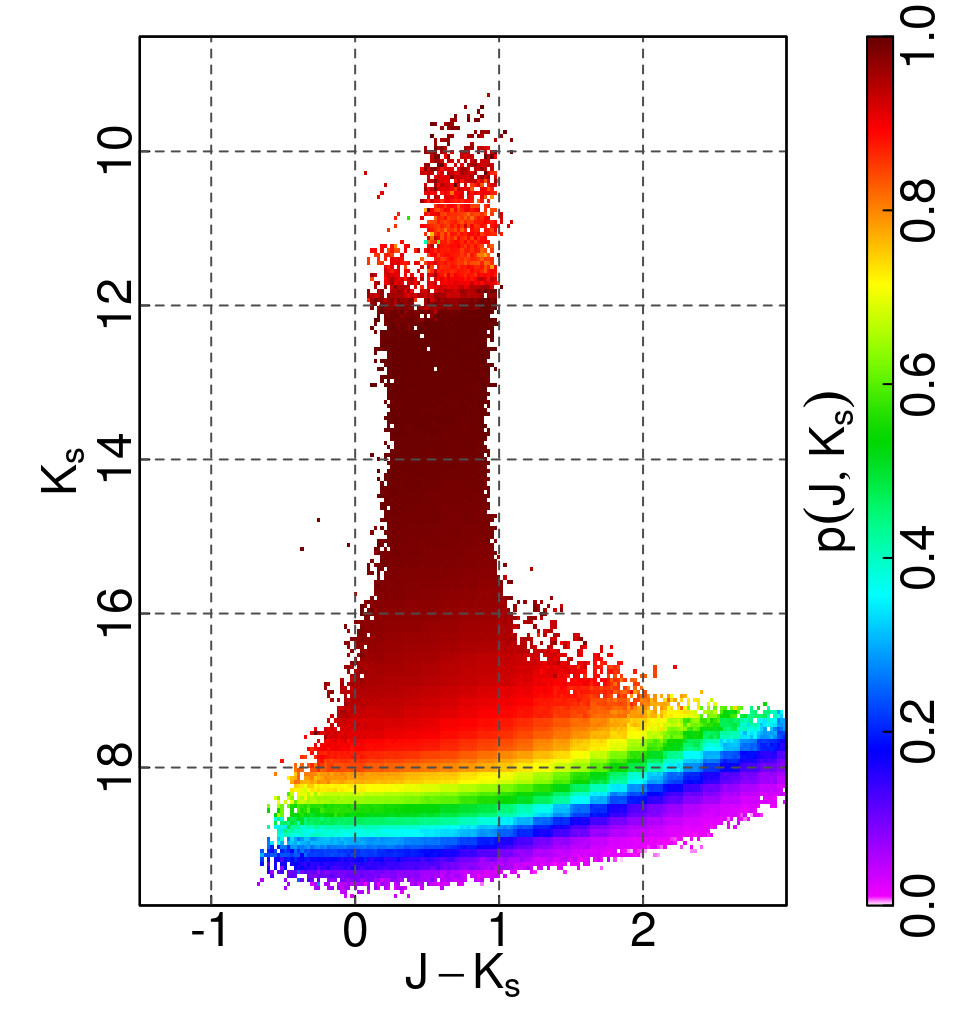}
\includegraphics[height=5 cm, trim={0 3.9cm 5.7cm 0}, clip]{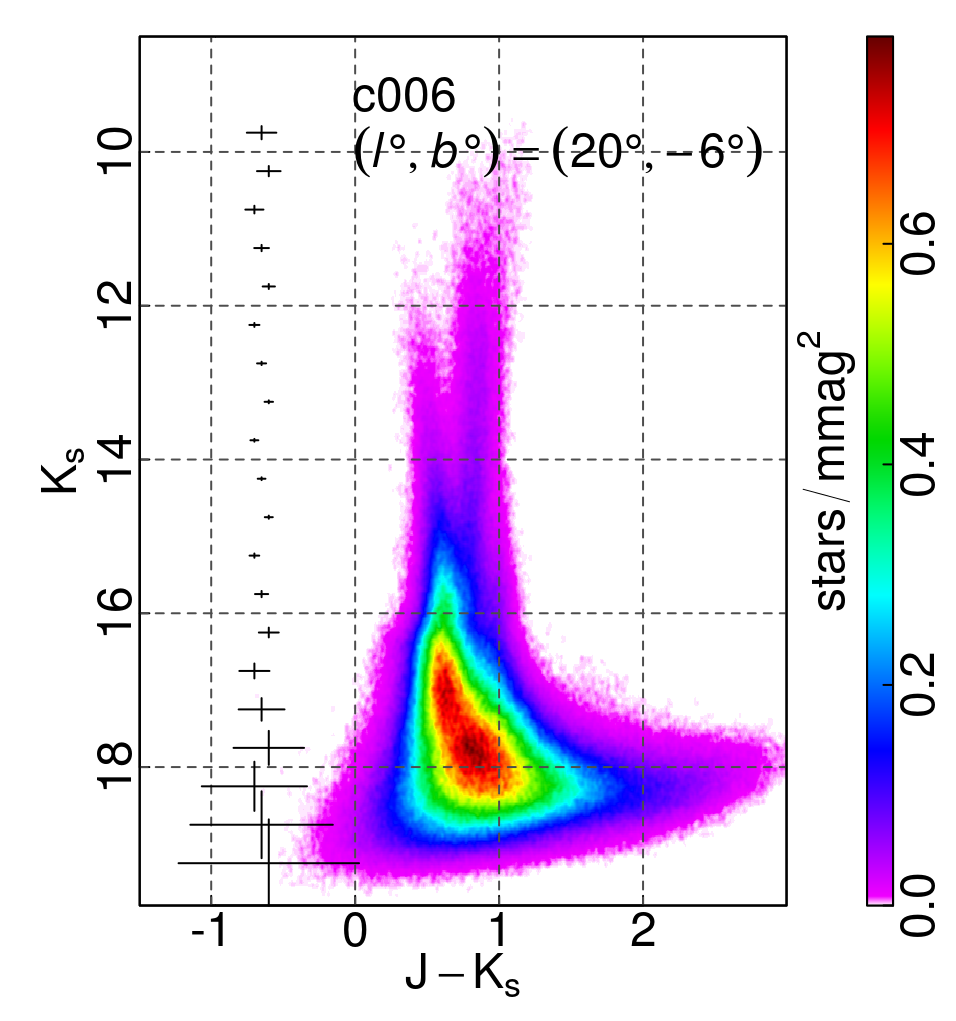}
\includegraphics[height=5 cm, trim={4.8cm 3.9cm 0 0}, clip]{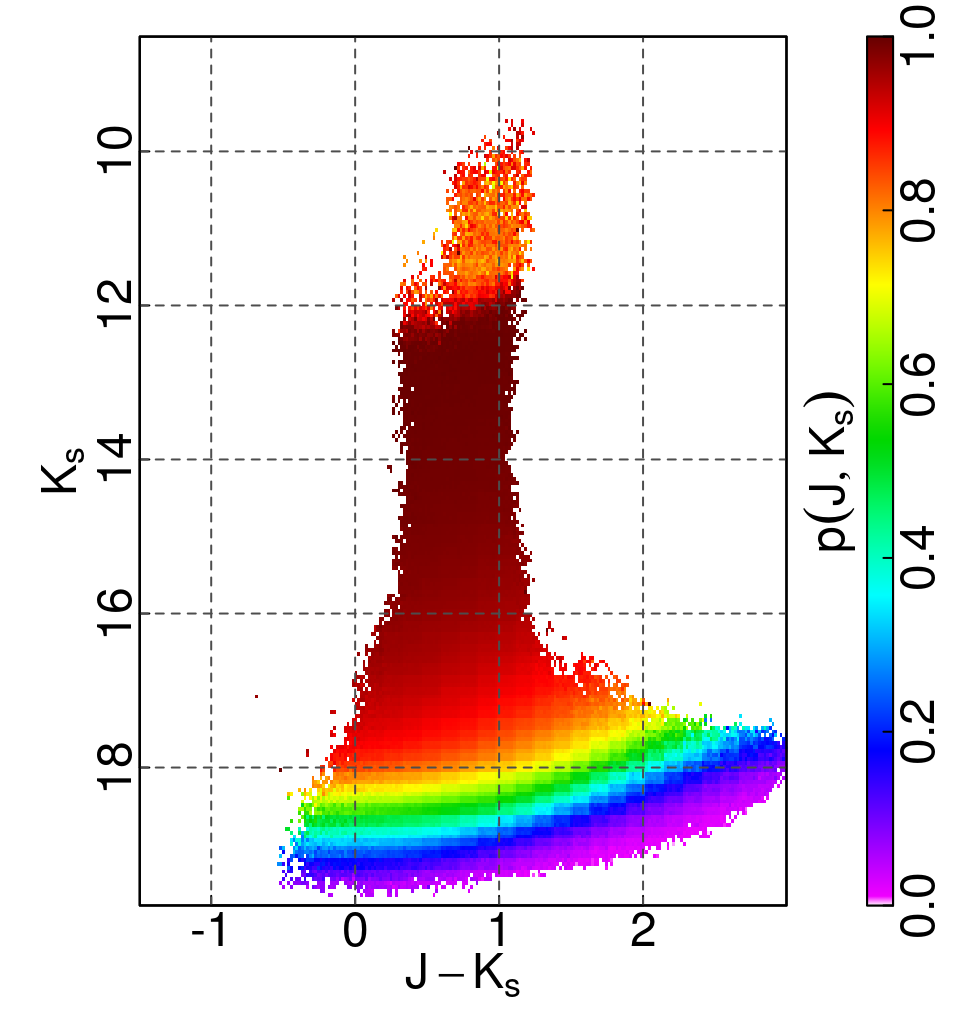}
\includegraphics[height=5.6 cm, trim={0 0 5.7cm 0}, clip]{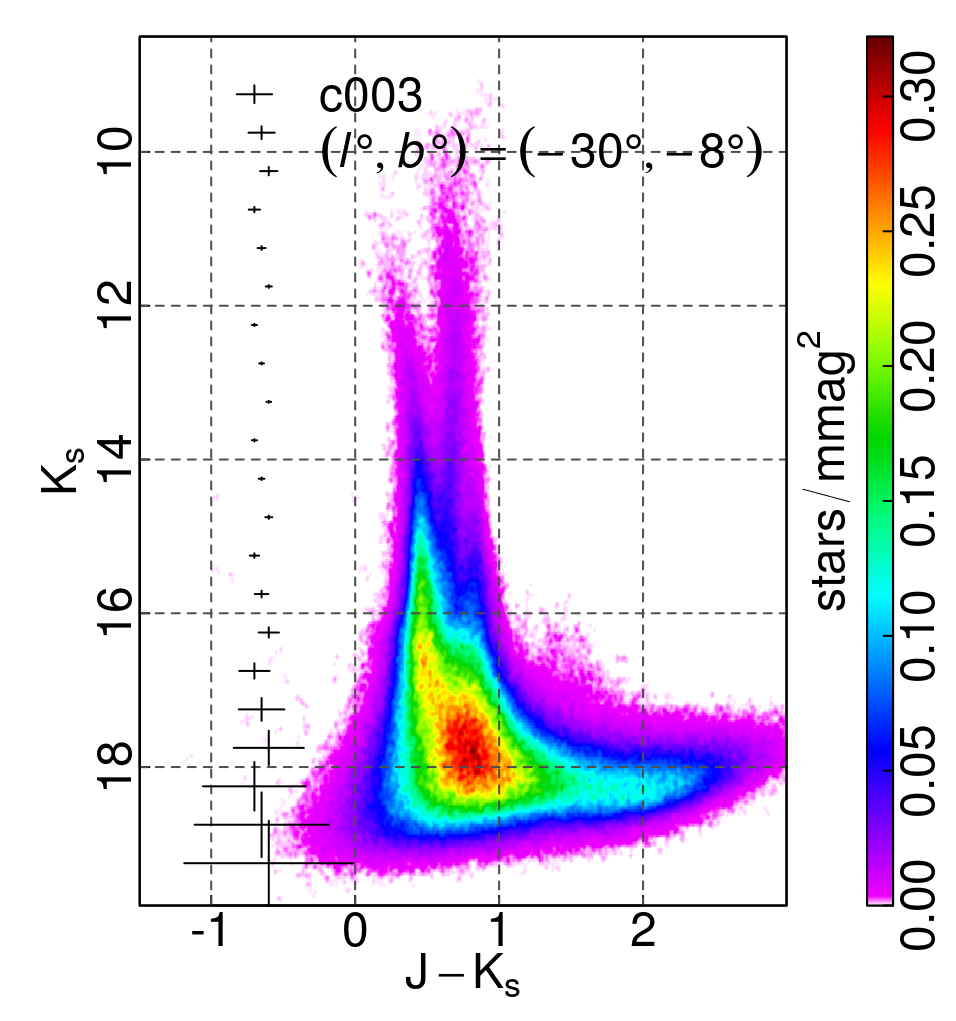}
\includegraphics[height=5.6 cm, trim={4.8cm 0 0 0}, clip]{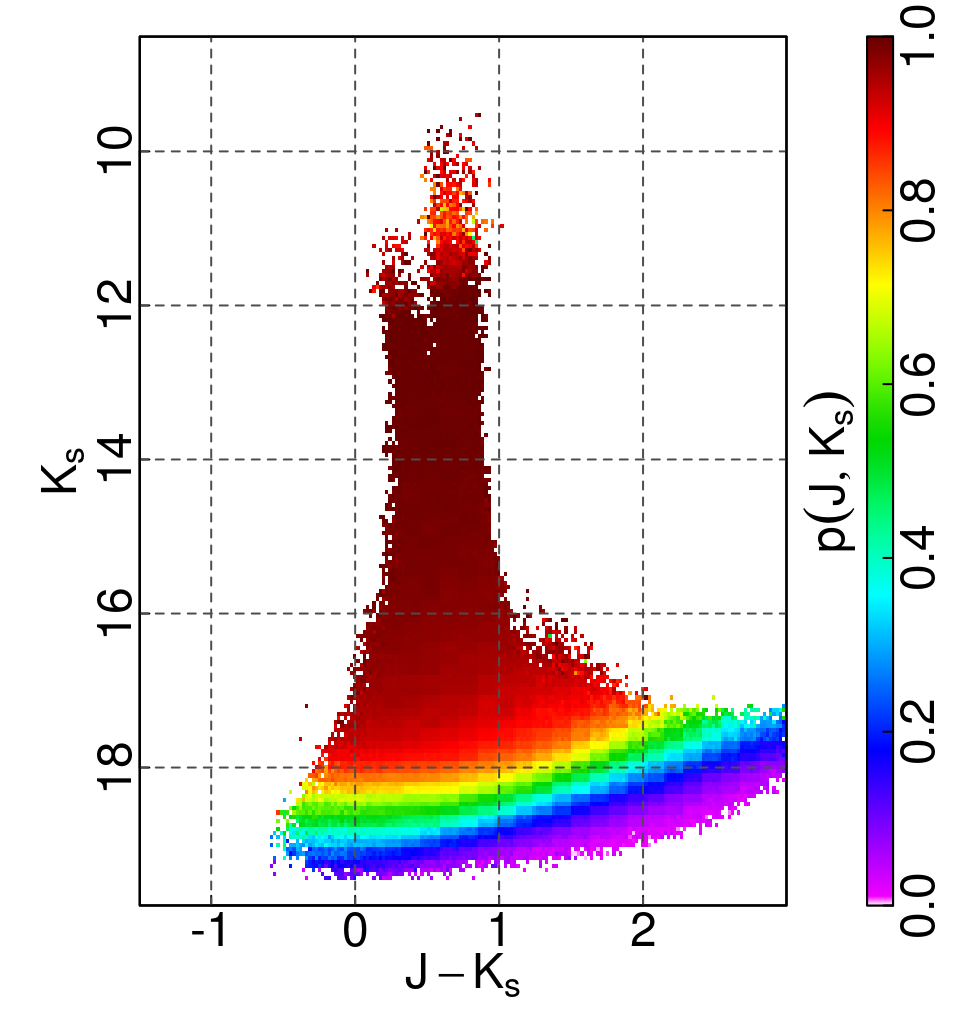}
\includegraphics[height=5.6 cm, trim={0 0 5.7cm 0}, clip]{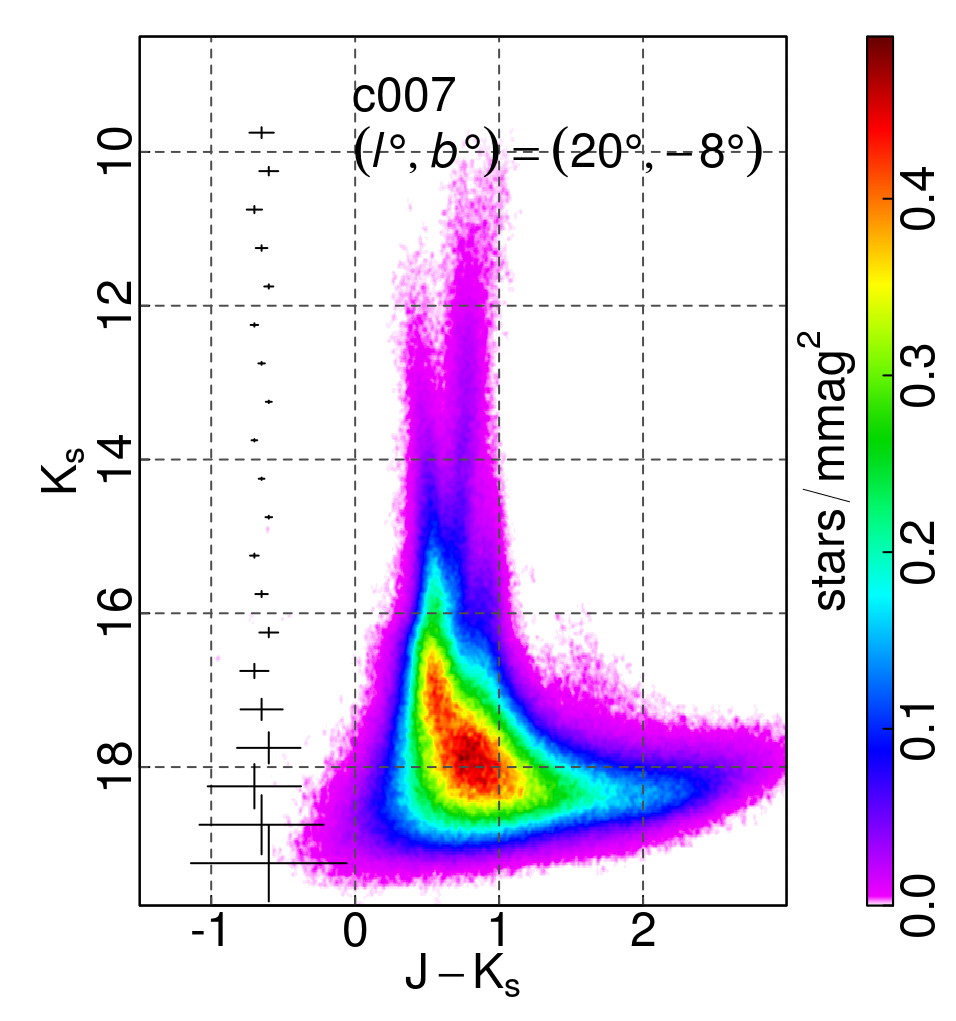}
\includegraphics[height=5.6 cm, trim={4.8cm 0 0 0}, clip]{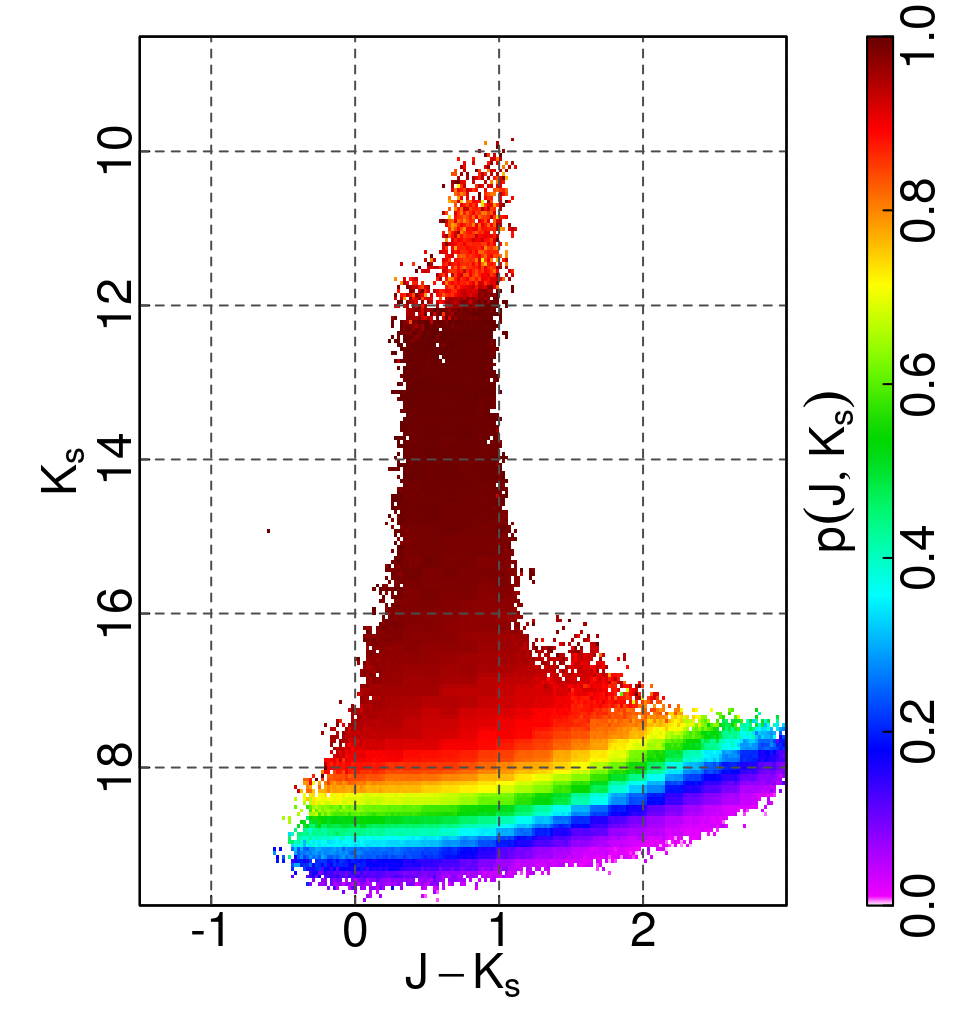}
   \caption{Observed CMDs of the 8 disk\--control fields (left panels) and the corresponding photometric completeness map (right panels).}
      \label{fig:DiskObsCMDs}
\end{figure*}

For each targeted field, the simulations yield the fraction $p = p(\mathrm{J},\mathrm{K_s})$ defined as the number of injected stars recovered with $|m^{in}-m^{rec}| < 0.75$ mag, to the total number of injected stars per color\--magnitude bin. Therefore, $p = p(\mathrm{J},\mathrm{K_s})$ is the probability of observing a given star with J and $\mathrm{K_s}$ magnitudes (i.e. magnitude and color) within a specified bin, hence providing the completeness of the derived catalogs.
As expected, $p$ decreases for fainter magnitudes and redder colors, however we also found that it changes on a detector by detector basis, with absolute differences of the order of $\Delta p = 0.1$ around the faint magnitude end of the CMD.
In the case of the less crowded fields ($b \lesssim -6^\circ$) the simulations show that the photometric catalogs of the observed bulge and disk fields are more than 50$\%$ complete above J$\sim$18.5 and $\mathrm{K_s}\sim$18 (see middle panel of Fig.\,\ref{fig:BulgeObsCMDs}, and right panels of Fig.\,\ref{fig:DiskObsCMDs}). The 50$\%$ completeness line moves to brighter magnitudes the closer the field is to the galactic plane.
 
 \section{The Observed Bulge Color\--Magnitude Diagram}
 \label{sec:cmd}
 
 The observed $(\mathrm{J}-\mathrm{K_s}, \mathrm{K_s})$ CMD of the selected VVV field (b249) is shown in the left panel of Fig.\,\ref{fig:BulgeObsCMDs}. 
When using standard point\--scatter plots, the large number of detected stars (see Tab.\,\ref{tab:obs-log}) saturates the plots for $\mathrm{K_s}>15$ preventing to distinguish different evolutionary sequences, such as the Sub Giant Branch (SGB) and the MS\--TO. Therefore we show the CMDs in Fig.\,\ref{fig:BulgeObsCMDs} as Hess density diagrams. 
 
The observed CMD shows a well populated RC at $\mathrm{K_s}\sim 13$ and $(\mathrm{J}-\mathrm{K_s})\sim0.8$; a prominent Red Giant Branch (RGB) easily traceable down to $\mathrm{K_s}\sim 16$; and the MS\--TO  at $\mathrm{K_s}\sim 17$ and $(\mathrm{J}-\mathrm{K_s})\sim0.5$. 
 On the other hand, the SGB is barely visible because it is heavily contaminated by the foreground disk population. Indeed, the two vertical blue sequences departing from the bulge MS\--TO and RC upwards correspond, respectively, to the foreground disk MS and its RC descendants.
 
 In the  magnitude range $10\leq \mathrm{K_s} \leq17.5$, the observed spread in all sampled evolutionary sequences is mostly due to the combination of metallicity and depth effect, whereas in the fainter range the contribution of the photometric errors becomes predominant. 
In b249 the differential reddening is not dramatically large ($\Delta E(J-K_s)\sim0.047~\mathrm{mag}$, see Tab.~\ref{tab:obs-log}), in fact when we correct for the extinction by using the VVV\--based extinction map from \citet{oscar12} the dereddened CMD  shows very similar magnitude spread (e.g. compare left\-- and right\--most panels in Fig.\,\ref{fig:BulgeObsCMDs}).

\begin{figure}
	\centering
	\includegraphics[width= 7.5cm]{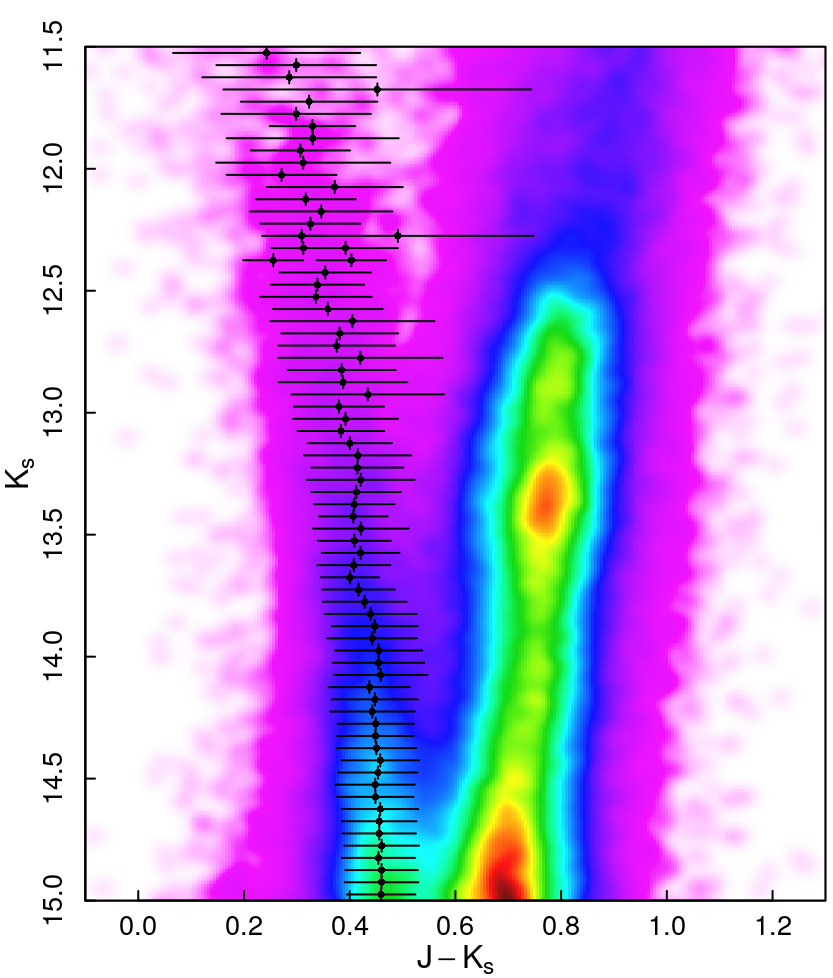}
	\caption{Color coded Hess density diagram of b249 compared to the gaussian young MS profile (black dots). Horizontal and vertical crosses refer respectively to the estimated sequence width and bin size.}
	\label{fig_ymsprofile}
\end{figure}

Finally, as expected the characteristic double RC signature of the well known X\--shape bulge structure is detected at $\mathrm{K_s}_0 = 12.75$ and $13.27$ (see Figure~\ref{fig_ymsprofile} for a zoom\--in). Indeed, as demonstrated by several authors (see \S\ref{sec:intro}) the X\--shape structure is detected only in the outer bulge, at latitudes $|b| > 5.5^\circ$ and longitudes $|l|\leq4.5^\circ$.

 \section{Disk decontamination procedure}
 \label{sec:decont}

As shown in Fig.\,\ref{fig:BulgeObsCMDs}, the MS of the disk hits the bulge CMD exactly on top of its MS-TO, hence preventing any reliable age determination. Therefore, prior to any attempt to age\--dating the bulge population from its observed CMD, a special care must be paid to remove the contribution of the intervening disk population along the line of sight. 
This can be done  either by using proper motions \citep[see][]{clarkson08PPM, clarkson11BS, kuijken+02, bernard+18}, or statistically with a disk control\--field  \citep[e.g.][]{zoccali+03, valenti+13}.  Both methods require some assumptions, and therefore have their own pros and cons. 

In principle, the proper motions selection yields more accurate decontamination providing that bulge and disk populations have clear distinct kinematics along the line of sight of interest, which is not always the case. 
In fact, as evident from Figure 4 of \citet{bernard+18}, along different lines of sight the proper motions of disk and bulge largely overlap, therefore there is no simple clear cut in $\mu_l$ and/or $\mu_b$ that would help differentiate disk from bulge; although a filter in proper motions can still be applied, it would inevitably end with an equally problematic cross\--contamination between the populations.
In addition, from the observations point of view the determination of proper motions is very time consuming. The need for precise astrometric measurements and long time-baselines for error reduction effectively limits this type of observations to expensive and accurate observatories such as HST, which has a small field of view (i.e. $\lesssim 3 \times 3$ sq. arcmin). Finally, it is worth mentioning that the proper motions provided by Gaia (DR2) do not allow a proper {\it cleaning} of the CMDs.
Being severely limited by both crowding and reddening already at $b=-6^\circ$, the Gaia photometry is too shallow and does not fully sample the old MS\--TO. 

On the other hand, the statistical approach best suits the case of large surveyed areas such as that presented here, but it relies on the assumption that the adopted control\--field is representative of the disk population along the bulge line of sight. In this respect, the selection of the disk control\--field is crucial and must take into account that the contamination of the bulge CMDs from foreground disk stars strongly depends on latitude.

\subsection{Comparable Populations}
\label{sec:decont_compops}

Figure\,\ref{fig:DiskObsCMDs} shows the derived Hess density diagrams in the $(\mathrm{K_s}, \mathrm{J}-\mathrm{K_s})$ plane of the 8 disk control\--fields studied here, together with the corresponding completeness map obtained from the artificial stars experiment (see \S\ref{sec:data}). 

In the blue side of the diagram for $(\mathrm{J}-\mathrm{K_s})\lesssim0.5$, one can easily identify the very well populated MS, while the redder vertical sequence corresponds to the evolved RC population. 
At fixed latitude, the sampled disk population in the fields at longitude $+20^\circ$ and $-30^\circ$ has fairly similar CMD, with the major difference being the overall reddening, which appears to be smaller at $-30^\circ$. As a consequence, the color of the MS and RC stars is generally bluer that what is observed at $+20^\circ$. 

As expected from an exponential disk density profile, at fixed longitude, the number of detected stars increases in fields closer to the Galactic plane, whereas at given latitude, the fields at $+20^\circ$ are systematically more populous than their counterpart $-30^\circ$  (see Tab.\,\ref{tab:obs-log}).

The first step of the decontamination process is the selection of the field that best represents the disk population observed along the bulge line of sight. 
This is done by comparing the bright portion of the CMDs ($\mathrm{K_s}\lesssim15$) in the bulge and disk fields. 
Specifically, we trace the profile of the young disk MS and the RGB by means of a series of gaussian fits to their $(\mathrm{J}-\mathrm{K_s})$ color distribution per $\mathrm{K_s}$ magnitude bin (see Fig.~\ref{fig_ymsprofile}). 
It is worth mentioning that these sequences lie on the bright and most complete part of the CMD, and as such have the smallest errors in general.

Shifts in color and magnitude along the reddening vector \citep[see][]{nishiyama+09,oscar12} are applied to all disk control\--fields to match the profile of the young blue MS of each control\--field with that observed in the b249 field. The shifts applied here are carried over for all subsequent analyses.
In doing this, we are ignoring distance distribution differences between the bulge and control fields. However because of the lack of secure and recognizable standard distance features in the disk sequences, there is no way to properly account for them. 
We compare the relative differences between the profiled young MS in b249 and in all disk fields, and select the one with the lowest dispersion of the residuals, which are in turn defined by the separation of the profiled young MS of the control field, interpolated onto the observed bulge counterpart. Following this procedure, the disk field c002 located approximately at the same latitude of the target bulge region turns out to be the most appropriate for decontamination purposes.

After choosing the best candidate control field, the next important step is to make sure that its dispersion in color and magnitude due to the combination of systematics and photometric uncertainties (see \S\ref{sec:data}) is comparable to the one observed in the bulge field.
We have in principle two kinds of uncertainties. 
The first is related solely to the PSF fitting procedure, and to the counts of the individual star profiles in an image taken with the photometric filter $M$: $\sigma_{M}$ (i.e. $\mathrm{M=J, K_s}$). This is our photometric error, which is tied to the shape and brightness of the individual stars. 
The second one comes from the measured dispersion in the completeness experiments: How artificial stars with similar injected magnitude $m^{in}$ disperse into randomly different recovered magnitudes $m^{rec}$. 
We call this  $\Sigma_M = \Sigma_M^{field}(M)$, which is mostly a function of $\mathrm{J, K_s}$, unique to each {\it field} (i.e. detector and image), and likely comprising the systematics in our data. 
To estimate the latter, for each detector we use a $4^{th}$\--degree polynomial to fit the $\mathrm{K_s}$\--binned $m_{in}$ vs. $\mathrm{MAD}(m^{in}-m^{rec})$ profile, where MAD is the adjusted median absolute deviation (i.e.  1-MAD is equivalent to 1-$\sigma$ from a normal distribution).

We need to take into account differences in observing conditions and crowding for the observed disk and bulge fields, which result in differences in PSF fitting and the associated error profiles. Given these differences between images as well as when considering individual detectors, we must be sure that the $i$-th star in the control field catalog, with magnitude $M^i$, has a similar error than an equivalent star would have in the bulge catalog. 
To do this, we define a value $\varsigma^i_M$ for the $i$-th star in the control field, as the maximum between its  photometric error, completeness dispersion $\Sigma_M^\mathrm{c002}(M^i)$ and the corresponding $\Sigma_M^\mathrm{b249}(M^i)$ from the observed bulge field.

   \begin{equation}
	\varsigma_M^i = \mathrm{max}\left(\sigma^i_M,\Sigma_M^\mathrm{c002}(M^i),\Sigma_M^\mathrm{b249}(M^i)\right)
   \label{eq:sigmad}
   \end{equation}

Similarly, as mentioned in \S~\ref{sec:data}, each field and detector have their own completeness $p^{field}(\mathrm{J},\mathrm{K_s})$, therefore a further step to guarantee a proper statistical subtraction is to correct the control field by its completeness and then apply the completeness of the observed bulge field. 
In practice, this can be achieved by assigning a weight $\omega^i$ to the $i$-th star in the control catalog, defined as:

   \begin{equation}
	\omega^i = p^\mathrm{b249}(\mathrm{J^i},\mathrm{K_s^i})/p^\mathrm{c002}(\mathrm{J^i},\mathrm{K_s^i})
   \label{eq:omegai}
   \end{equation}

Finally, we must calculate the bulge\--to\--disk normalization factor, which gives us the number of stars to be removed from b249 for each given disk star observed in c002. 
To do so, we select a region in the b249 CMD where one is likely to find only the disk population (i.e. $(J-K_s)\lesssim0.5$, and $K_s\lesssim16$), as previously done by \citet{zoccali+03, valenti+13}. This factor is a single scalar applied uniformly throughout the control CMD. 

\begin{figure}
	\centering
	\includegraphics[width= 9 cm]{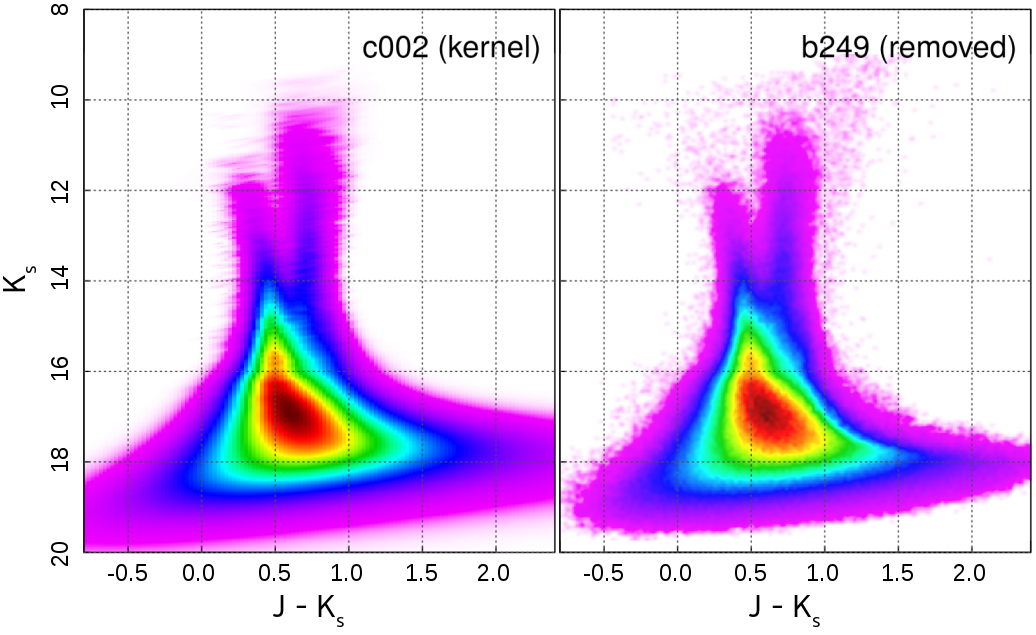}
	\caption{{\it Left}: Intensity kernel map for c002, as constructed from b249 dispersion. 
	{\it Right}:CMD of the stars removed from b249 by using the kernel map shown in the left panel. 
	}
	\label{fig:removed}
\end{figure} 

\subsection{Kernel approximation and subtraction}
\label{sec:decont_kernel}

To take into account the effects of the error bars and systematics in the bulge and disk CMDs, we adopted a bivariate gaussian kernel smoothing map. 
This is similar to modeling any given star in the [$\mathrm{K_s}$, $\mathrm{J}-\mathrm{K_s}$] plane as a bivariate normal distribution, whose centroid is just the color\--magnitude position of the star, and its covariance matrix constructed from the errors of J and $\mathrm{K_s}$. 
We then stack/add all the gaussians, and evaluate the result on a finer grid, so that now the integral of this yields the expected number of stars in any given region of the CMD.
As $\sigma_{\mathrm{K_s}}$ and $\sigma_\mathrm{J}$, we take the corresponding $\varsigma_M$ from (\ref{eq:sigmad}), and approaching the problem as if all stars had errors defined by this quantity.

Because the errors and dispersion are a function of the $\mathrm{J}$ and $\mathrm{K_s}$ magnitudes for any given star, there is no unique kernel valid for any given catalog. Therefore we divide the dataset in $\sigma_\mathrm{J}$-$\sigma_\mathrm{K_s}$ bins, calculate the kernel map for each, and add them together.
Once the kernel map is constructed, we scale it by using the bulge\--to\--disk normalization factor (see \S~\ref{sec:decont_compops}) in order to ensure that the expected number of stars in the kernel map and in the observed bulge CMD is the same in the defined disk\--only window.
After estimating the expected number of stars from the kernel map in small regular bins across the color\--magnitude space, we extract the corresponding number of stars by randomly picking the catalog entries from each bin.
We then count the number of stars within the defined window for the remaining stars and compare it with the original observed catalog in order to get a new scale factor (usually a small correction of the order of $10\%$) and repeat once more the subtraction. 

The remaining stars represent the effective bulge clean sample.
It is important to note that this procedure yields bulge\--like and disk\--like catalogs according to their CMDs only. 
In other words, if we had additional information (e.g. kinematics) that would allow us to clearly distinguish disk and bulge stars individually, and used that on either clean or removed catalogs, neither would show a pure disk or bulge population, but the resulting CMDs would look like as if they were, regardless of their true precedence.

\section{The bulge clean sample}
\label{sec:clean}

We performed the decontamination procedure described in \S~\ref{sec:decont} on the observed bulge field b249 by using c002 as disk control\--field.

In Figure~\ref{fig:removed} we show the kernel map of c002 (left panel), and the density map of the stars subtracted from b249 (right panel). The overall similarity of the two maps provides a first sanity check, demonstrating that the procedure worked as intended. 
Indeed, for $\mathrm K_s\lesssim19$ and $0\lesssim \mathrm J-\mathrm K_s \lesssim 1.2$ the maps are virtually indistinguishable, except for the 2 faintest magnitude bins where the kernel shows a wing extending red\--wards that is not reflected in the removed set. 
In this case, the wing extends towards rapidly declining completeness (i.e. see Fig.~\ref{fig:BulgeObsCMDs}), which is not taken into account in the kernel map but only considered in the weight of the entries of the control catalog. 

\begin{figure}
    \centering
    \includegraphics[width=\hsize]{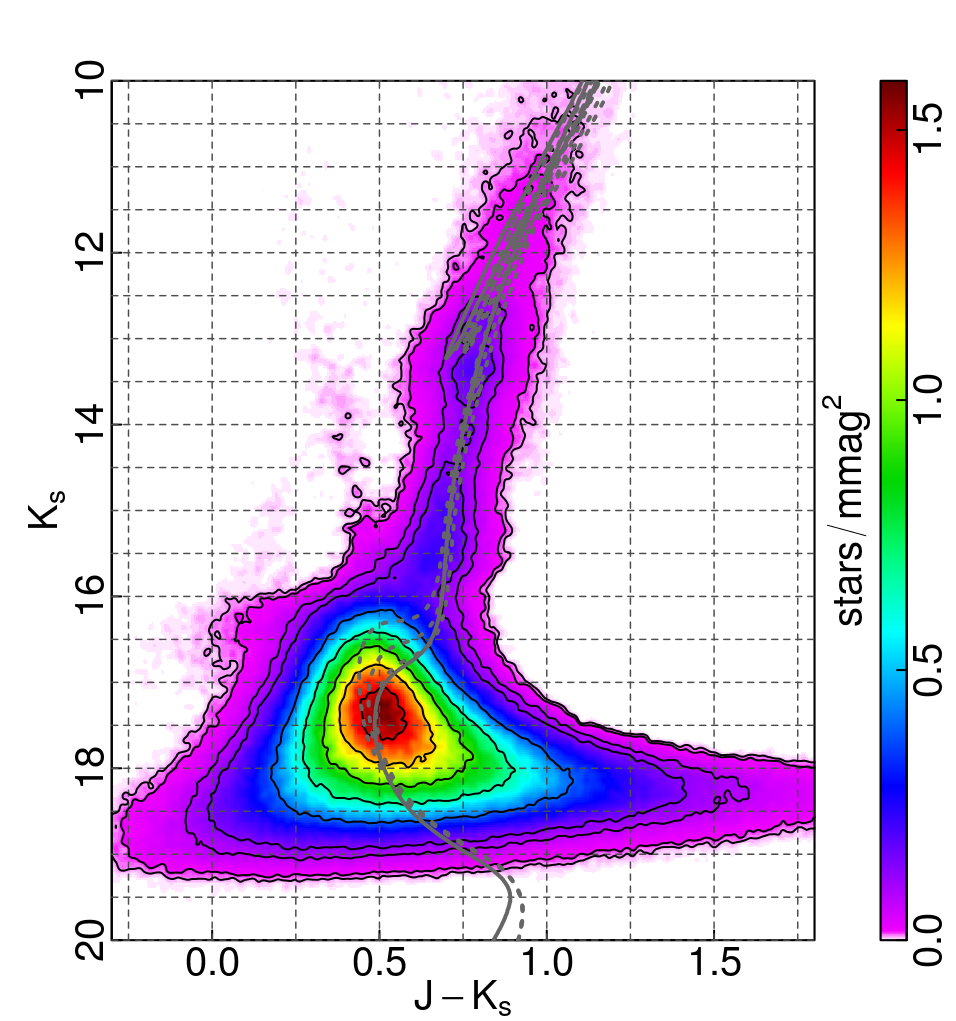}
    \caption{Hess diagram of the bulge b249 field as statistically decontaminated from the foreground disk population. Solid black contours are isodensity curves spaced by 1\%, 5\% and then from 1/12 to 11/12 of the maximum density, in steps of 1/6. Also plotted are BaSTI isochrones for ages (bluest to reddest) 5, 7.5, 10 and 11 Gyr, following the metal\--poor r\'{e}gime from GIBS for the oldest (grey lines), and metal\--rich for all others (dotted line). The isochrones have been shifted so that their RC and RGB loosely match the observed ones.}
    \label{fig:b249clean}
\end{figure} 

The final result of the decontamination procedure is presented in Fig.~\ref{fig:b249clean}, where we show the CMD of the bulge b249 clean sample, consisting of 1,654,603 stars. 
By removing the foreground disk population, one can now more easily recognize the bulge main evolutionary features: the double RC $\mathrm{K_s}=12.83, 13.35$ and  $(\mathrm{J-K_s})=0.79$), the RGB ($\mathrm K_s \lesssim 15.5, \mathrm{(J-K_s)} \gtrsim 0.5$), the SGB ($\mathrm 16.5 \gtrsim K_s\gtrsim 15.5$, and $\mathrm 0.4 \lesssim\mathrm{(J-K_s)}\lesssim 0.7$), and the hot spot near $\mathrm K_s \sim 17$ with a shape resembling a MS\--TO.
The halos silhouetting the subtracted young disk MS are outliers that could not be removed completely. However, their signal in the Hess density diagram is very weak (i.e. $\sim 4000$ stars, $<0.2\%$ of the maximum density value, and $\sim 11\%$ of the original observed in the same region), therefore they can be safely neglected. 
Finally, the blue feature near $\mathrm{(J-K_s)}\sim 0.1$ and $\mathrm K_s\sim 16.5$ is likely an artifact due to stars mostly located at a corner of detectors 2, 8 and 16. The diagnostic values (shape and $\chi^2$ from PSF\--fitting) of these artifact have different distributions to that of normal stars, but they still overlap, making an usual cut (e.g. in $\chi^2$) ineffective to clean them.  This feature was of course present also in the observed CMD (i.e. Fig.~\ref{fig:BulgeObsCMDs}), but since it is not real nor present in the control field, it became highlighted after the decontamination.

\section{Stellar ages determination}
\label{sec:age}

To determine the age of the bulge stellar population in the b249 field a simple isochrone fitting of the clean sample does not yield meaningful results because of the spread in color and magnitude induced by the observational effects around the expected MS-TO (see Fig.\,\ref{fig:b249clean}).
In addition, because we are studying a complex stellar population over a large surveyed area, the metallicity and large distance spread introduce further dispersion in color and magnitude that cannot be qualitatively taken into account by using simply isochrones. 
Instead, we resort to the comparison of the bulge clean sample with synthetic populations with known age and metallicity. 

The synthetic populations are obtained with the  IAC\--star code \citep{iacstar}, employing the BaSTI stellar evolution models \citep{abasti, bbasti, obasti, ubasti}, and by using the empirical IMF of  \citet{calamidaimf} with a 35\% binary fraction.
 
We adopt the spectroscopic MDF provided by the GIBS survey \citep{gibsIII}, and based on 213 RC stars observed in the b249 field. 
The observed b249 MDF is best represented by two gaussian components: the MR distribution peaks at  $[\mathrm{Fe/H}]_{MR} = +0.08 \pm 0.16~\mathrm{dex}$, while the MP at  $[\mathrm{Fe/H}]_{MP} = -0.35 \pm 0.17~\mathrm{dex}$. 
Moreover, to further constrain the synthetic populations, we adopt a 1:1.2 ratio of MP to MR, as suggested by \citet{gibsIII}, and an $\alpha$-elements enhancement of +0.31 for the MP population \citep[][and references therein]{argosIII, gibsII, mcwilliam16}. We note that this consideration towards differing [$\alpha$/Fe] is merely an approximation. A more careful approach should account for $[\mathrm{Fe/H}]$\--to\--$[\mathrm{\alpha/Fe}]$ relations for both metallicity r\'{e}gimes.

We also add the reddening to the synthetic population from a random draw, emulating the color excess distribution of the stars in the b249 field according to the extinction map of \citet{oscar12}. 
To account for the depth effect we use the split RC information of the observed field. We combine two identical synthetic populations, with a grey magnitude shift equal to the distance between the split RC gaussians, and add to each the corresponding grey gaussian spread, obtained from the quadratic difference between the gaussians fits of the RC in observed and synthetic CMDs. The combination is constructed so that the star count ratios between the RC gaussians in the synthetic CMD is equal to that in the observed CMD. Based on the RC distribution, which as we pointed out in \S\ref{sec:clean} includes two overdensities, the distance probed in the observed bulge field is approximately between 6 and 11\,kpc.

\begin{figure*}
	\ContinuedFloat*
	\centering
	\includegraphics[width=.43\hsize ,trim={0 0 4.8cm 0}, clip]{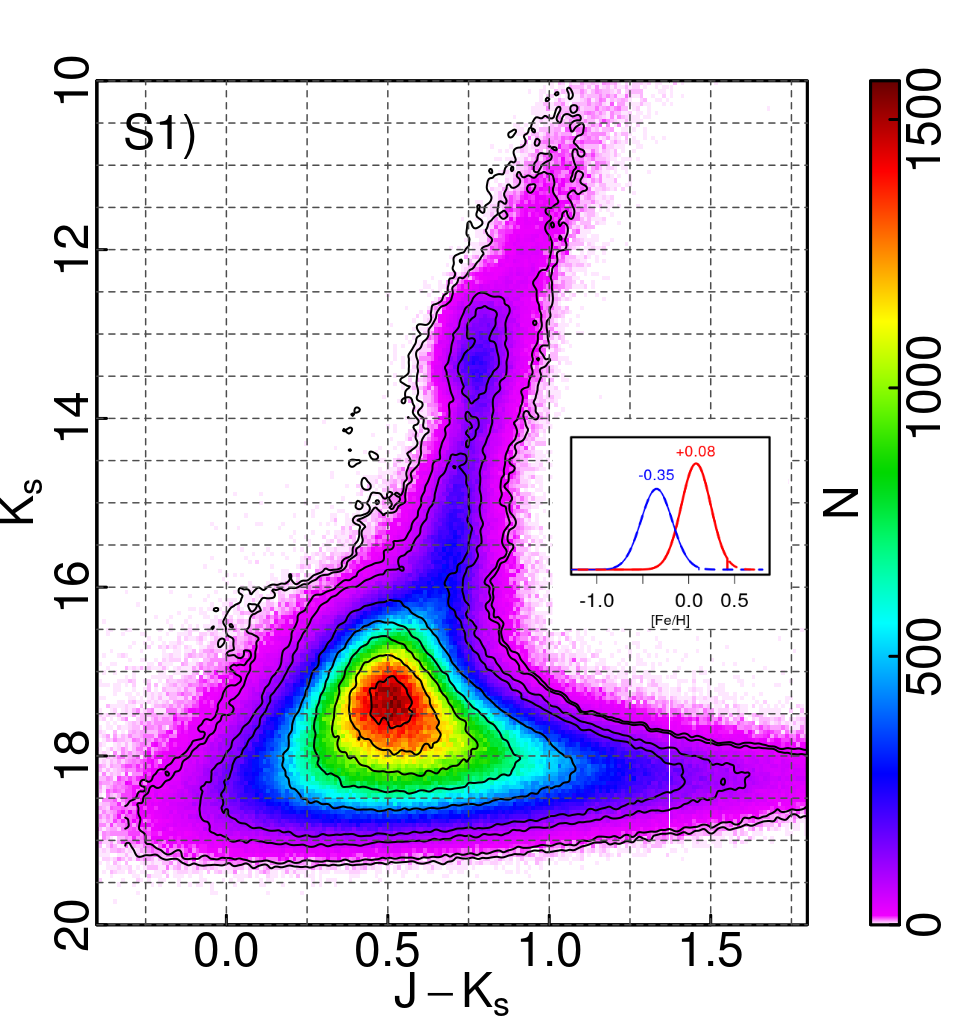}
	\includegraphics[width=.453\hsize, trim={3.25cm 0 0 0}, clip]{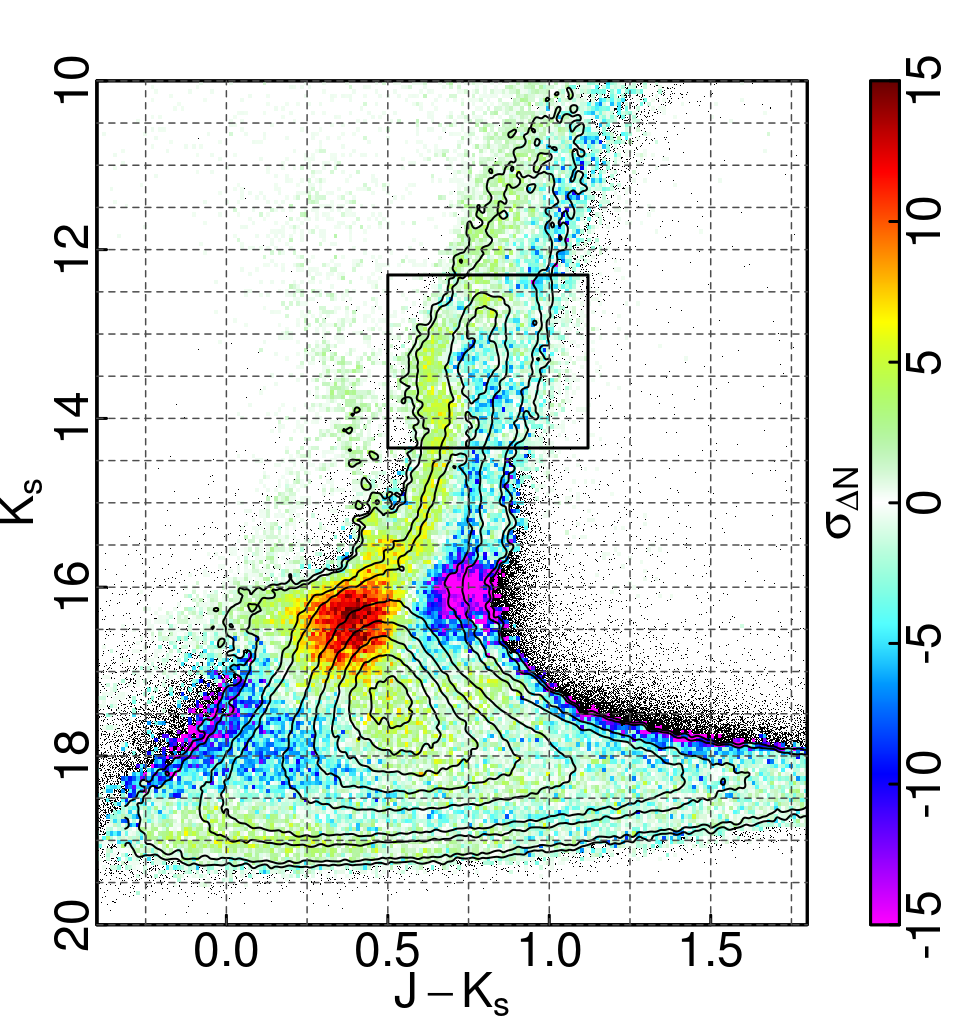}
	\includegraphics[width=.43\hsize ,trim={0 0 4.8cm 0}, clip]{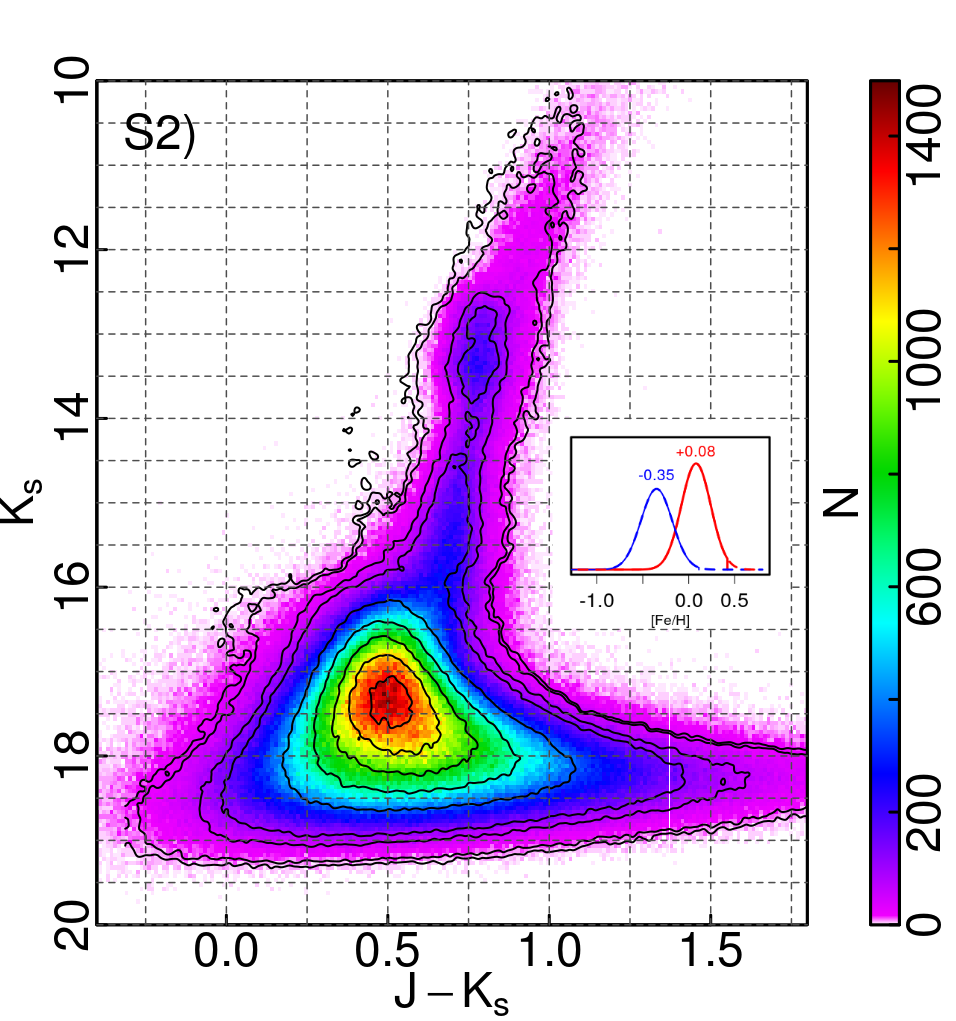}
	\includegraphics[width=.453\hsize, trim={3.25cm 0 0 0}, clip]{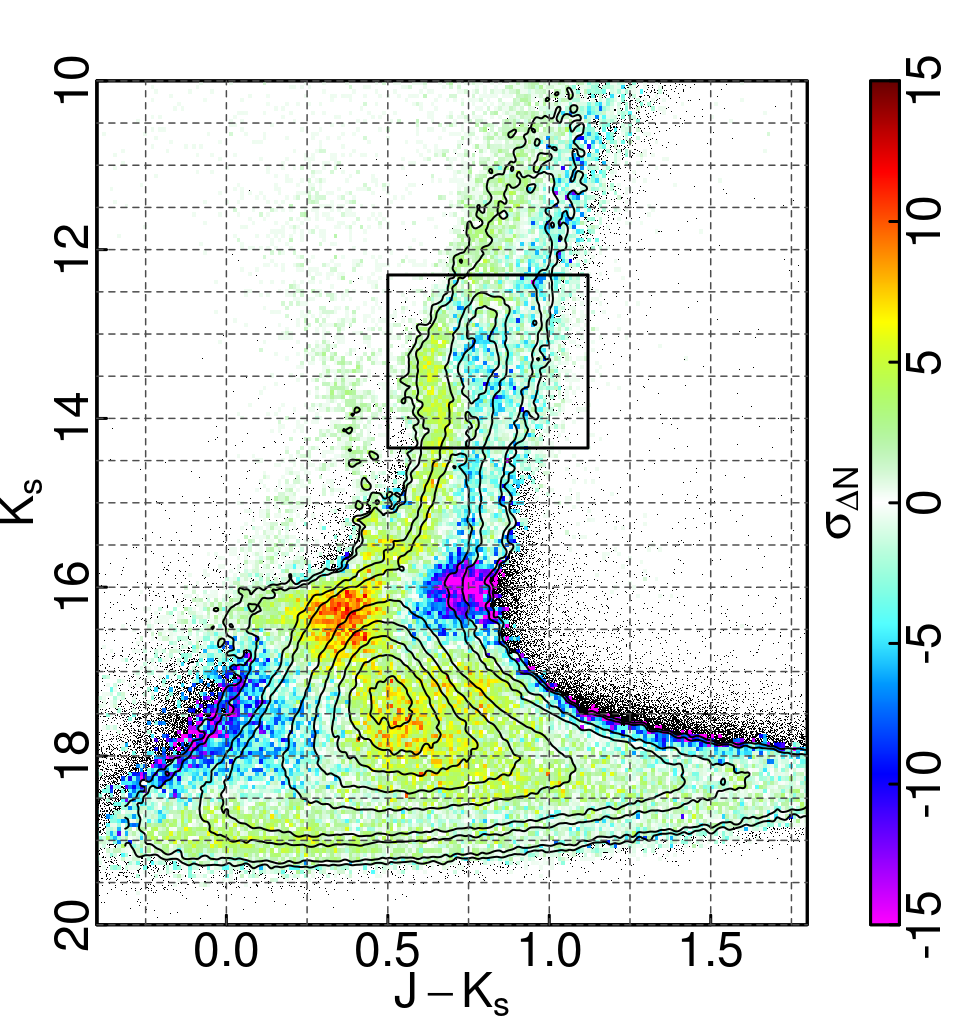}
	\caption{Left\--panels: synthetic CMDs corresponding to the six scenarios described in \S~\ref{sec:age} compared to the isodensity curves (solid black contours) of the  observed clean sample as derived in Fig.\,\ref{fig:b249clean}.The adopted MDF for all (but S4) scenarios is shown in the insets (see text).	
Right\--panels: residuals map providing the dispersion (e.g. the quality of the fit) between the synthetic and observed CMD. Significant mismatch between the observations and simulations corresponds to $|\sigma_{\Delta N}| \gtrsim7$. Regions of the CMD populated exclusively by simulated stars are marked with black dots (see text). The boxes on the right\--most panels mark the region of the CMD used to normalize the number of simulated stars to the observations.}
	\label{fig:b249simul}
\end{figure*} 

\begin{figure*}
	\ContinuedFloat
	\centering
	\includegraphics[width=.43\hsize ,trim={0 0 4.8cm 0}, clip]{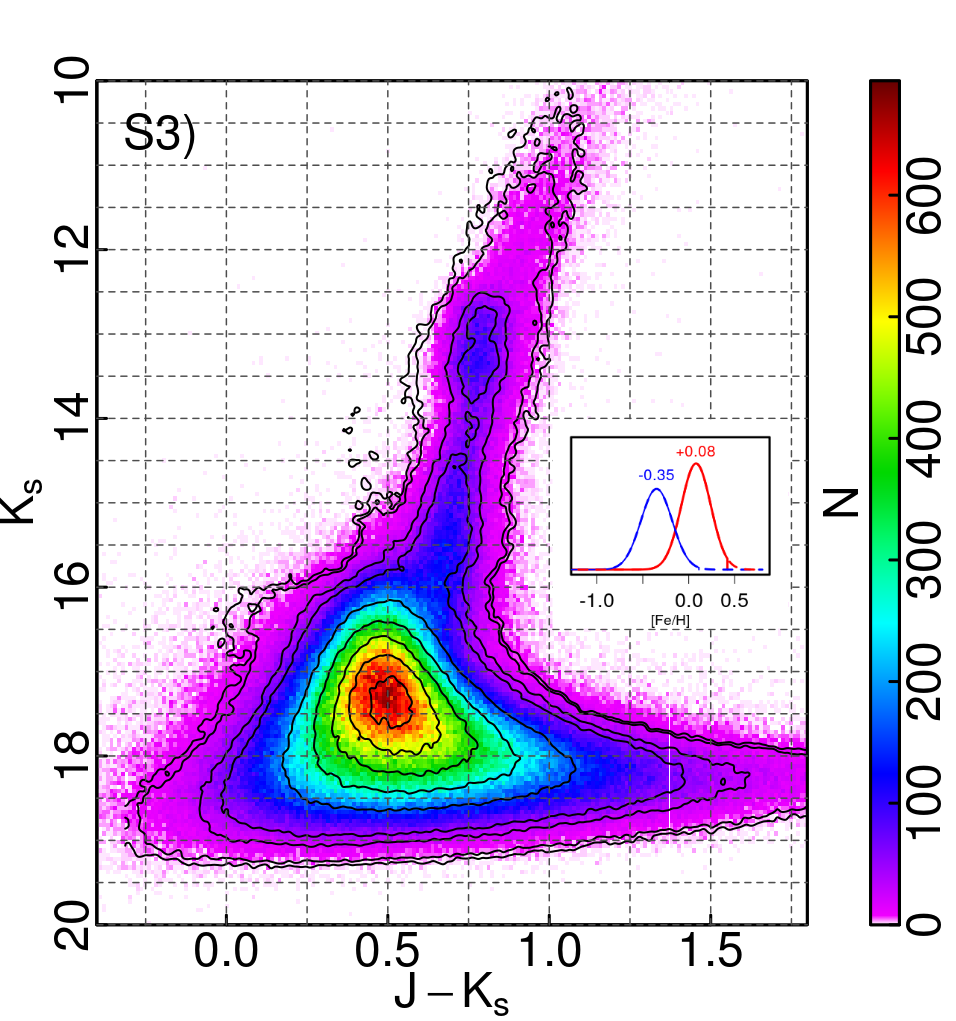}
	\includegraphics[width=.453\hsize, trim={3.25cm 0 0 0}, clip]{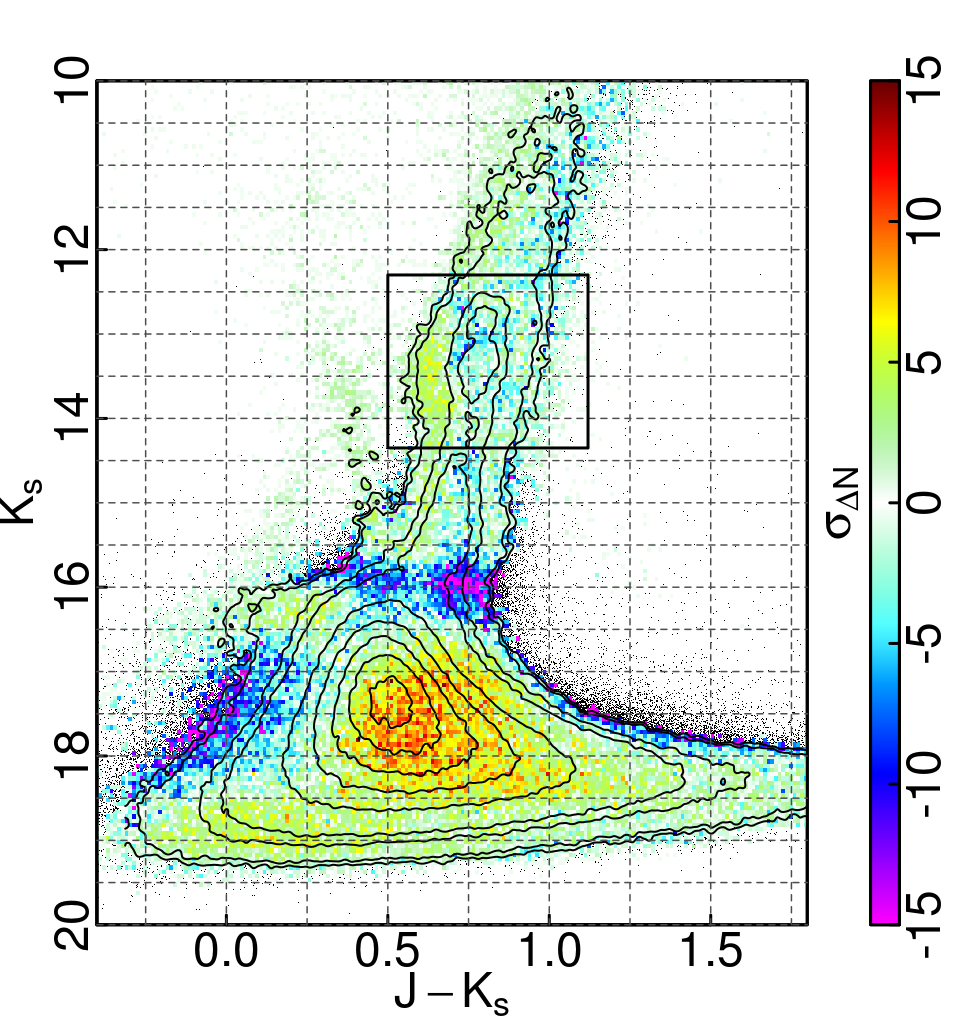}
	\includegraphics[width=.43\hsize ,trim={0 0 4.8cm 0}, clip]{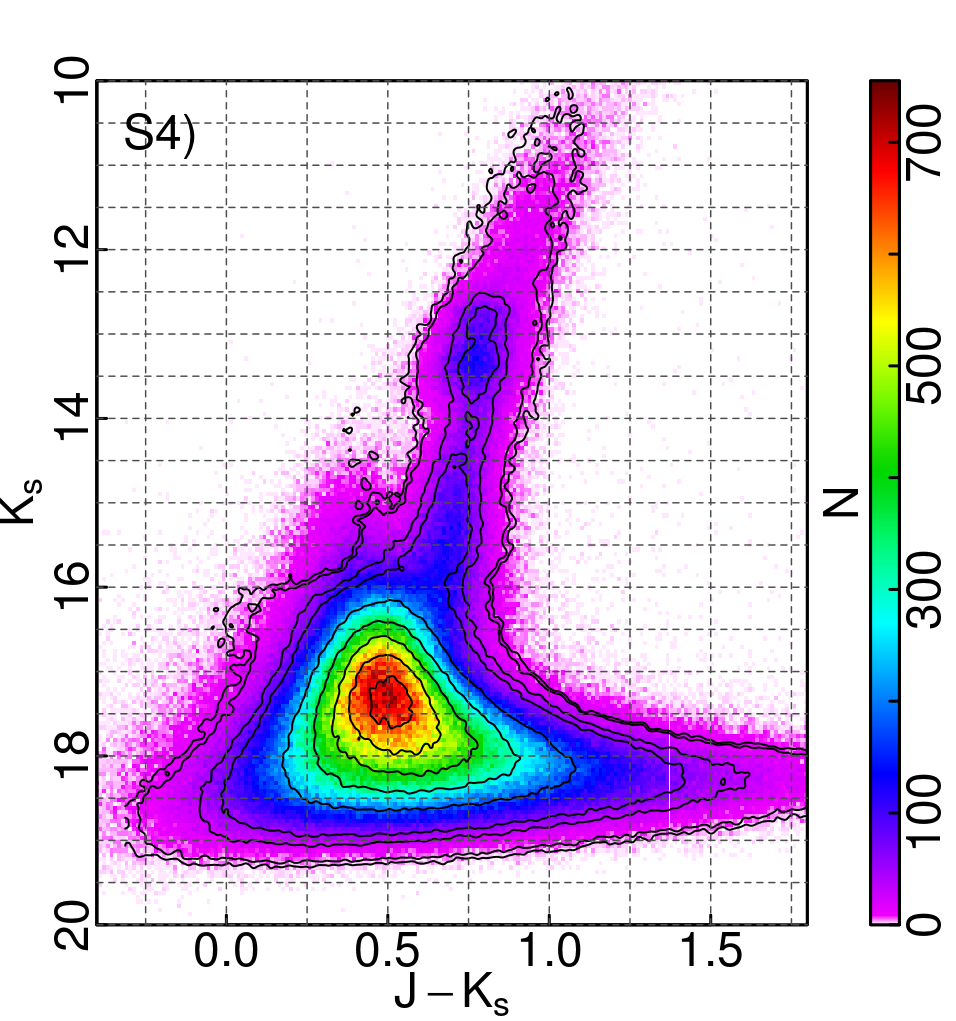}
	\includegraphics[width=.453\hsize, trim={3.25cm 0 0 0}, clip]{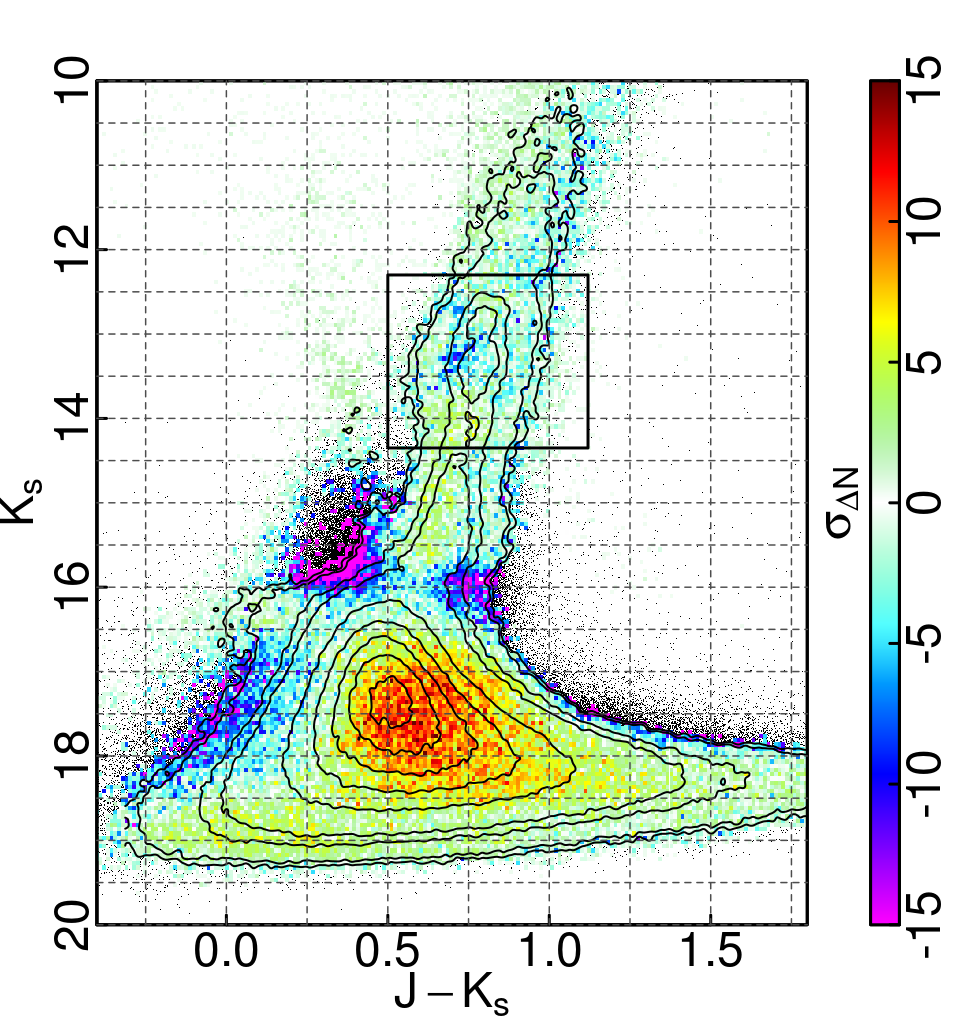}
	\caption{(continued)}
	\label{fig:b249simul_2}
\end{figure*}

\begin{figure*}
	\ContinuedFloat
	\centering
	\includegraphics[width=.43\hsize ,trim={0 0 4.8cm 0}, clip]{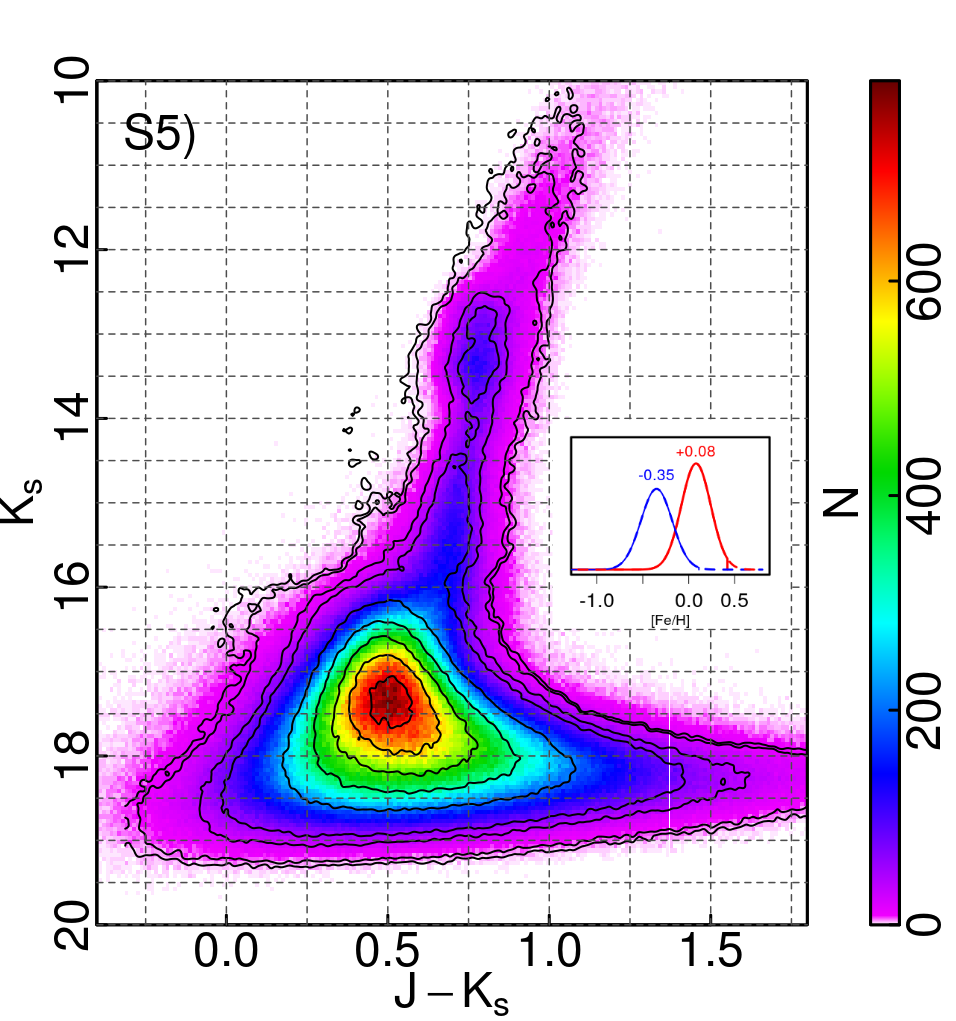}
	\includegraphics[width=.453\hsize, trim={3.25cm 0 0 0}, clip]{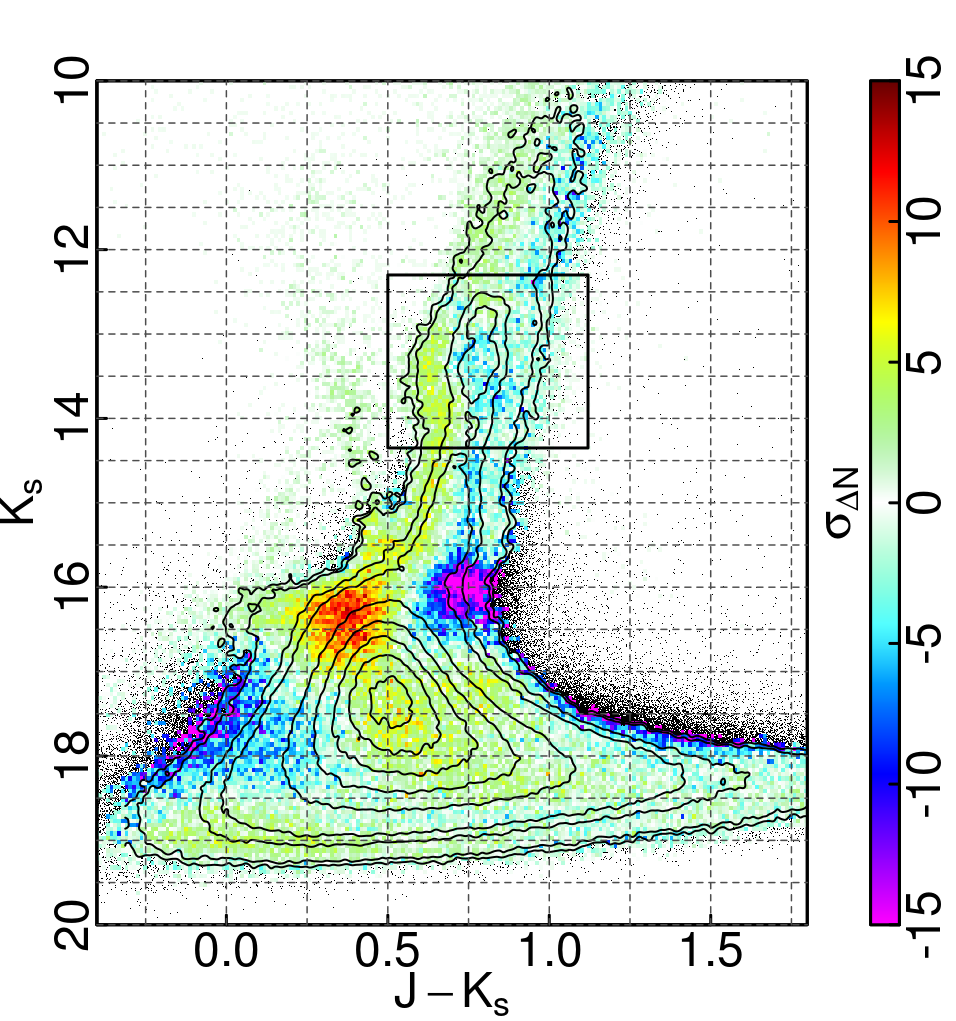}
	\includegraphics[width=.43\hsize ,trim={0 0 4.8cm 0}, clip]{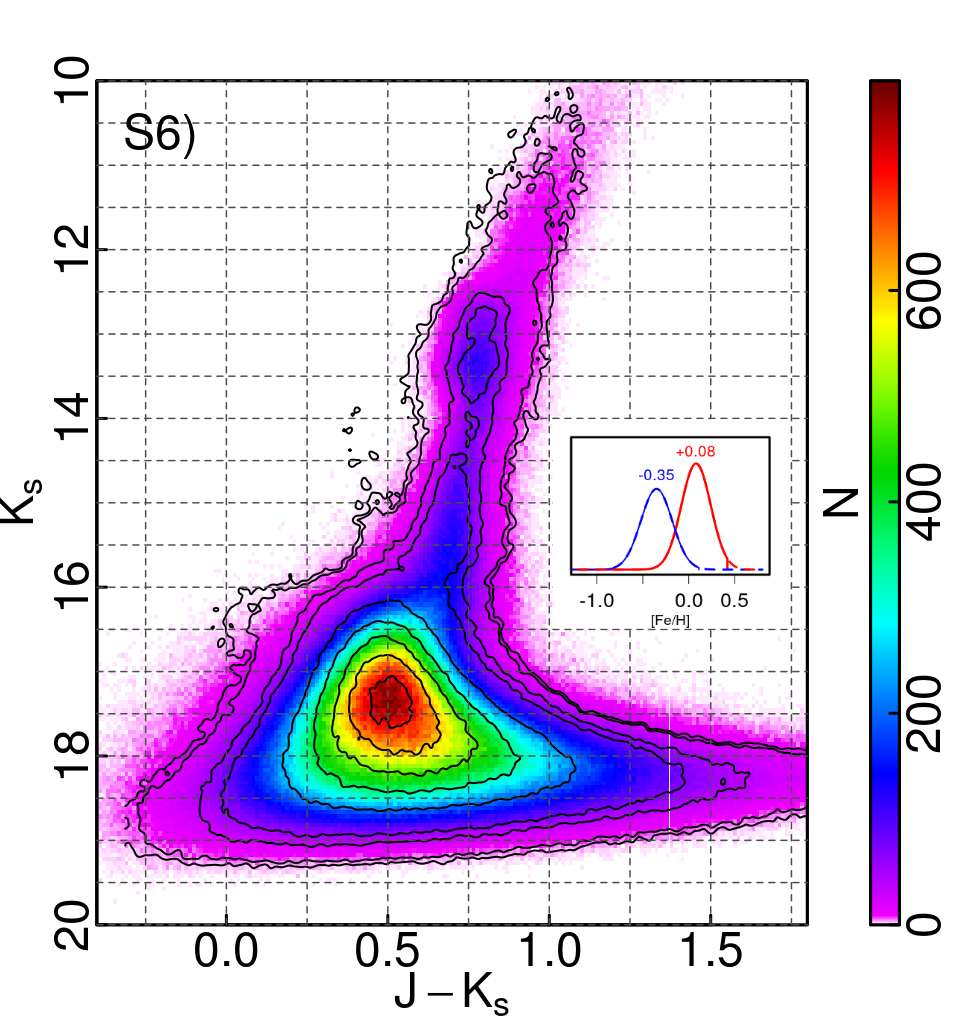}
	\includegraphics[width=.453\hsize, trim={3.25cm 0 0 0}, clip]{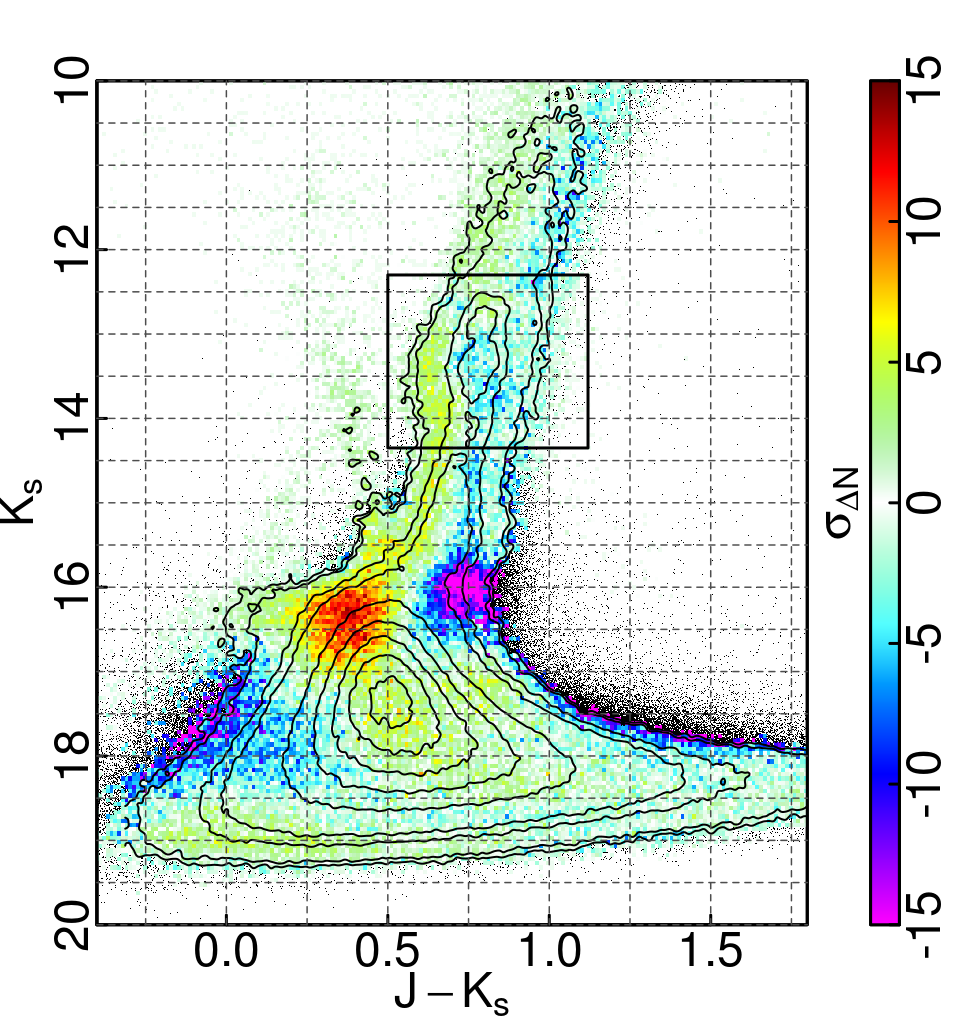}
	\caption{(continued)}
	\label{fig:b249simul_3}
\end{figure*} 


Theoretically, once the metallicity, $\alpha$\--enhancement, distance, and reddening of the synthetic population are constrained, as well as a given binary fraction, reddening law and IMF are assumed, the age becomes the only free parameter. 
Hence, in principle we could let the age varies until we find the best match to the observed clean sample.
In practice, however, the situation is further complicated by the dispersion due to the observational uncertainties (i.e. photometric errors and completeness). 
Any comparison between synthetic and observed CMDs aimed at providing reliable age estimates must take into account the observational effects. 
This is best done by dispersing the synthetic populations according to a new completeness map specifically tailored for this purpose. 
We remind here that the artificial stars experiment described in \S~\ref{sec:data} has provided a completeness (i.e. $p = p(\mathrm{J},\mathrm{K_s})$) for all the stars in the observed catalog only. 
Now, if we want the age of the synthetic population to vary freely we need to quantify the completeness and the distribution of the differences $\mathrm{ (m^{in} - m^{rec})}$ also for some regions that are not populated in the observed CMD.
We achieve that by adding $\sim \! 1,000,000$ artificial stars, per detector, distributed around the same spacial grid as in \S~\ref{sec:data}, but covering a specific region of the CMD ($\mathrm 20 \lesssim K_s\lesssim 9$, and $\mathrm -0.2 \lesssim\mathrm{(J-K_s)}\lesssim 1.4$) where we know model stars lie, and then constructing a new completeness map{\footnote{To run these CPU intensive simulations we used resources of the Computational Center for Particle and Astrophysics (C2PAP). https://wiki.tum.de/display/c2pap2018/C2PAP}} following the same procedure outlined in \S~\ref{sec:data}.

According to the models, the observed RC shape suggests that the MP component must be quite old, between 11 and 12 Gyr, therefore in our simulations we allow to vary the age of the MR component only. 
Note that this is also in line with all previous studies (see \S~\ref{sec:intro}) that found the MP bulge stellar population to be consistently older than $\sim$10\,Gyr. 
Indeed, if a significant fraction of a younger population is present then it is likely to be more metal\--rich than the older one.

Although our simulations explored a fairly large age range $\mathrm{Age\,(Gyr)} \lesssim 13.5$, in the present paper we present and discuss only a sub\--sample, which can be considered as the most representative of the entire set of simulations.

Specifically, for the age of the two components we considered the following six different scenarios:

\begin{itemize}
\item {\bf S1}: $\mathrm{MP=(11\pm0.3)\,Gyr}$, $\mathrm{MR=(10\pm0.3)\,Gyr}$  
\item {\bf S2}: $\mathrm{MP=(11\pm0.3)\,Gyr}$ , $\mathrm{MR=(7.5\pm0.3)\,Gyr}$  
\item {\bf S3}: $\mathrm{MP=(11\pm0.3)\,Gyr}$, $\mathrm{MR=(5\pm0.3)\,Gyr}$  
\item {\bf S4}: age\--metallicity distribution from B17 as obtained from their Figure 14 (i.e. from individual stars).
\item {\bf S5}: $\mathrm{MP=(11\pm0.3)\,Gyr}$, \\
$\mathrm{MR=40\% (10\pm0.3)\,Gyr + 60\% (7.5\pm0.3) \,Gyr}$  
\item {\bf S6}: $\mathrm{MP=(11\pm0.3)\,Gyr}$, \\
$\mathrm{MR=60\% (10\pm0.3)\,Gyr + 40\% (7.5\pm0.3) \,Gyr}$  
\end{itemize}

The ages are always flat distributions, and for all simulations but S4 we adopt the MDF spectroscopically derived by the GIBS survey that is best represented by two gaussians for the MR and MP components. 
In addition, as mentioned before, we adopt the MR/MP ratio found in GIBS implying that the number of RC stars in the MR and MP domains has the ratio 1.2:1.  
The results of the simulations for the six scenarios are shown in Fig.\,\ref{fig:b249simul}, where in the left panels we compare the Hess density diagrams of the synthetic population with the isodensity contours corresponding to the clean observed sample as derived in Fig.\,\ref{fig:b249clean}. 

To assess the quality of the match between simulations and observations and deriving the best\--fit model, we create for each simulation the corresponding residuals map (see right\--hand panels in Fig.\,\ref{fig:b249simul}).
Specifically, the noise of the observed CMD has been first modeled with a bootstrap approach. 
We repeatedly ($\sim\,10,000$ realizations) take a random resample of the complete observed catalogs, and save the residual between the subtraction of the original and resampled CMD density for each iteration. 
This effectively produces a residual dispersion per color\--magnitude bin of the observed CMD, which is used to measure any significant deviation when comparing the synthetic populations (i.e. subtraction between the synthetic and observed CMD). 
Following this approach, the perfect match between simulation and observation should have a residuals map that is characterized only by noise, hence by the lack of any structures. We have decided to favor a map comparison approach over a single\--digit quantification of a goodness\--of\--fit (e.g. $\chi^2$), because we expect not only the quantification of the discrepancies between observed and modeled datasets to be relevant, but also where in the CMD these discrepancies arise. For instance, a concentrated high\--$\sigma$ difference at $\mathrm{K_s} < 16$ can be as important as a wider spread lower\--$\sigma$ difference at the $\mathrm{K_s} > 17$ level.

The residual maps shown in Fig.\,\ref{fig:b249simul} have been scaled to the same color\--magnitude window and identical $\sigma$ ranges, hence they can be directly compared against each other. 
Because the structures built upon the noise appear only above $| \sigma | \approx 6$, to be conservative the signal of a given feature in the map is considered significant if $| \sigma | \gtrsim 7$ (i.e. above green and below cyan according to the adopted color code in  Fig.\,\ref{fig:b249simul}). In addition, we have adopted the simple convention where a positive $\sigma_{\Delta N}$ value implies excess observed stars with respect to the simulation, while negative values imply the opposite.

From Fig.\,\ref{fig:b249simul} we can rule out the possible presence in the bulge of a significant fraction of intermediate\--young populations (i.e. $\lesssim\,5$ Gyr). 
In fact, the residuals map of S3 and S4 shows a significant ($\sigma_{\Delta N} \gtrsim 8$) mismatch around the expected MS\--TO (the roundish structure that coincides with the highest isodensity contours). In addition, the S4 scenario predicts the presence of a large number of stars brighter than $\mathrm {K_s} \sim16$ and $0.25\lesssim \mathrm{J-K_s}\lesssim 0.5$. which are not observed (shown as small black dots in the residuals map of S4 in Fig.\,\ref{fig:b249simul}). Additionally, S4 has a slightly better fit for both RC and blue RGB, although this likely comes from the wider MDF used for this simulation.

On the other hand, the case of a purely very old bulge, represented by simulation S1, provides a good match of the region around the MS\--TO, but significantly ($\sigma_{\Delta N} \gtrsim 12$) underestimates the number of stars observed at $16\lesssim \mathrm {K_s} \lesssim16.5$ and $0.25\lesssim \mathrm{J-K_s}\lesssim 0.5$ (red spot in the S1 residuals map in Fig.~\ref{fig:b249simul}). 
The region at the base of the RGB, $\mathrm{(J-K_s)}\sim 0.75$ and  $\mathrm {K_s} \sim16$, is also not well reproduced, but the presence of this mismatch in all simulations suggests that it is an artifact caused by the decontamination procedure.  
In fact, it correspond to the bright end of the local M dwarf distribution clearly observed in the disk fields as the reddest vertical sequence around $\mathrm ({J-K_s}) \sim1$. 
As such, even if its intensity varies across the simulations, we refrain from taking it into account for the selection of the best\--fit model.
 
When considering a slightly younger age (i.e. $\sim\,7.5$\,Gyr) for the MR component, as in S2, S5 or S6, the overall match between the synthetic and observed population further improves, while the region around the MS\--TO is still well matched. 
Indeed, the deficit of synthetic stars present in S1 is considerably reduced, up to a factor of $\sim$\,2 in the S2 case.  
This would make S2 the best\--fitting scenario, however we should stress that in all simulations presented here we have ignored the presence of BSS even though they have been observed in bulge fields \citep{clarkson11BS}. 
To qualitatively constrain their position in the $\mathrm{[K_s, J-K_s]}$ plane, we have used the near\--IR CMD of the bulge cluster NGC6624 from \citet{saracino+16}, which includes BSS identified from UV and optical (Ferraro {\it private communication}). 
When accounting for the difference in distance and reddening between the cluster and the b249 field, we found that the BSS are approximately located in the region where S1, S2, S5 and S6 show a deficit of simulated stars with respect to the observed ones. 
Therefore, which simulation among scenarios S1, S2, S5 and S6 best fits the observed CMD, depends upon the number of BSS present in the field.

According to \citet{clarkson11BS}, the bulge field BSS frequency ($\mathrm{S_{BSS}}$), defined as the number of BSS scaled to the number of RC stars, is between 0.3 and 1.23, while for the clusters \citet{frf+03} found $0.1\lesssim \mathrm{S_{BSS}}\lesssim 1$. However, based on a photometric study of a sample of globular clusters and dwarf spheroidals, \citet{santana+13,santana+16} found that the number of BSS grows almost linearly with the total stellar mass of the system, therefore the BSS frequency in dwarf spheroidals is much higher than in clusters. In addition, when considering that the dynamical state of the bulge is expected to be more similar to that of dwarf spheroidals rather than clusters, it is plausible to believe that the bulge BSS frequency provided by \citet{clarkson11BS} could be an underestimation.

For the case of the b249 field, $\mathrm{S_{BSS}}$=1.23 would imply the presence of $\sim 32,000$ BSS.
Of course, to properly take into account the BSS in the simulation we would need to have robust information on their density distribution per color\--magnitude bin, which at the moment is still lacking. 
In principle, this can be obtained by using a large and statistically robust sample of observed BSS (Ferraro et al. {\it in prep}). 
Here we just stress that taking into account a population of $\sim 32,000$ BSS uniformly distributed within the region defined by the cluster NGC 6624, would have the effect of removing completely the mismatch between the observed and simulated CMD for the S2.
On the other hand, higher BSS frequency values  $\mathrm{S_{BSS}}$=1.66, 2.82, 3.03 need to be assumed to remove the stars deficit highlighted in the residuals map of scenarios S5, S6 and S1, respectively
Finally, the S3 and S4 scenarios are incompatible with any value of BSS frequency.

\section{Discussion and conclusions}
\label{sec:end}

We have used VVV images of the field b249 to perform new PSF\--fitting photometry, allowing to construct accurate and deep CMD that samples most of the evolutionary sequences, from the RGB down to $\sim \! 2$\, mag below the old MS\--TO.
By using a disk control field, a statistical decontamination procedure that accounts for relative completeness, extinction, photometric and systematic errors as well as relative population fractions, has been applied to extract a bulge  clean bulge sample. The corresponding CMD has been used to assess the stellar ages.
Due to the complexity of the system, instead of simple isochrone fitting, we have recurred to comprehensive simulated observations of composite populations exploring different scenarios in terms of stellar metallicity and ages.

When the MDF spectroscopically derived by GIBS is used, the best fit to the data is obtained with a composite population with ages in between 7.5\,Gyr and 11\,Gyr (i.e. scenarios S2, S5 and S6), therefore arguing for a formation scenario on a relatively short time scale ($\lesssim$\,5\,Gyr) in which the bulk of the bar/bulge stellar population was completely formed already at least $\sim$\,7.5\,Gyr ago.
However, although only qualitatively, we have also shown that when taking into account the presence of BSS in the bulge, the region of the CMD where one would expect to see intermediate\--young age stars (i.e. $\leq5$\,Gyr) gets populated.

In particular, we noted that the best fit to the data can be also obtained with a purely very old complex population (i.e. 10\,Gyr and 11\,Gyr as in S1 scenario) when assuming a BSS frequency of $\sim \!3$, thus suggesting an even early formation for the whole bulge stellar population.
This result would be in excellent agreement with the very recent study by \citet{renzini+18} based on very deep HST CMDs of four fields located along the bulge minor axis and $-4\lesssim b^\circ \lesssim -2$. By using a combination of UV, optical and near\--IR filters they have photometrically tagged all bulge field stars, and compared the luminosity function of the most MR and MP with simulated old and intermediate\--age population. The authors found that MR and MP populations appear essentially coeval and consistent with a $\sim$10\,Gyr old population.

Because BSS mimic a rejuvenated population due to the mass transfer, ignoring their presence could partially be one of the reasons that led previous studies to advocate the evidence of very extended star formation in the bulge \citep[e.g.][]{bernard+18}. 
Therefore, including BSS in the simulation of synthetic CMDs could potentially reduce, or even remove, the tension between different studies based on the photometric approach.   

On the other hand, the present study still does not reconcile the discrepancy between the photometric and spectroscopic age measurements.
In fact when we allow for the presence of a significant fraction (i.e. >\,20\%) of intermediate\--young stellar population (i.e. scenarios S3 and S4) the synthetic CMD does not provide a reasonably good fit of the observations. 
It should be stressed here that the comparison between simulations and observations has been performed on the properties of the entire CMD, not only in terms of color spread of the MS\--TO, as done in some previous studies \citep[e.g][]{haywood+16}. 
Stars as young as 1\,Gyr up to $\sim \! 3$\,Gyr occupy a region in the CMD that is not populated by the observations. 
Because our observed bulge sample is statistically very robust ($> \! 1.6\times10^6$ stars) due to the a large surveyed area ($\sim1.8\,\mathrm{deg}^2$), even a small fraction (i.e. $\gtrsim$\,10\%) of such young component, if present, would have been detected in the observed CMD. We note that the field b249 has only partial overlap with the surveyed area by \cite{bensby+17}, thus we cannot directly contradict the results derived by them. However, for both results to be in agreement, a steep age gradient favoring younger stars towards the Galactic center would be necessary (i.e. effective at about a degree closer to the center from b249). Even though N\--body simulations may predict such gradient \citep[e.g.][]{ness+14,debattista+17}, the number of stars in \citet{bensby+17} is insufficient to categorically establish the presence of a vertical gradient, specially at the outskirts of their surveyed area.

Finally, the method presented here will be applied to other VVV tiles in order to explore possible age distribution variations across the bulge. At the same time, we will compare the results with more sophisticated simulations using genetic algorithms that will ultimately lead to the reconstruction of the full star formation history.

\begin{acknowledgements}
FS, EV and MR acknowledge the support by the DFG Cluster of Excellence "Origin and Structure of the Universe". The simulations have been carried out on the computing facilities of the Computational Center for Particle and Astrophysics (C2PAP).
SC acknowledges support from Premiale INAF ”MITIC” and has been supported by INFN (Iniziativa specifica TAsP).
The authors thank Francesco R. Ferraro for the very useful discussion on BSS and for providing advanced information of the color and magnitude distribution in the bulge cluster.
Support for MZ and  DM  is  provided by  the  BASAL  CATA  Center  for Astrophysics  and  
Associated Technologies through grant PFB-06, and the Ministry for the Economy, 
Development, and Tourism's Programa Iniciativa  Cient\'\i fica Milenio through grant  IC120009, 
awarded to Millenium Institute of Astrophysics (MAS). 
EV and MZ acknowledge support from FONDECYT Regular 1150345.
DM acknowledges support from FONDECYT Regular 117012.
SH, SC and ES acknowledge grants 309290 form IAC and AYA2013-42781P from the Ministry of Economy and Competitiveness of Spain.

\end{acknowledgements}
\bibliographystyle{aa}
\bibliography{myrefs}

\end{document}